# Three-Dimensionality in the flow of an elastically mounted circular cylinder with two-degree-of-freedom vortex-induced-vibrations

Mayank Verma[1], Ashoke De[1,2 a)]

[1]*Department of Aerospace Engineering, Indian Institute of Technology Kanpur, Kanpur, 208016, India.*

[2]*Department of Sustainable Energy Engineering, Indian Institute of Technology Kanpur, Kanpur, 208016, India.*

[a)] *Corresponding author: ashoke@iitk.ac.in*

The study numerically investigates the three-dimensionality in the flow and two-degree-of-freedom (2-DOF) vortex-induced-vibrations (VIV) characteristics of an elastically mounted circular cylinder. The cylinder is allowed to vibrate in both streamwise and transverse directions. A low value of mass-ratio ($m^* = 2.546$) with the zero-damping coefficient ($\zeta = 0$) is taken for the simulations. The primary aim is to understand the vortex shedding behind the cylinder and the transition characteristics of the wake-flow from two-dimensional (2D) to three-dimensional (3D). The Reynolds number (Re) is varied from 150 (fully 2D flow) to 1000 (fully 3D flow), which lies inside the laminar range. The reduced velocity ($U_r$) is varied $2 \leq U_r \leq 17$ which covers all three major VIV branches (Initial Branch (IB), Upper Branch (UB), and the Lower Branch (LB)). The oscillating cylinder sweeps the figure-eight trajectory. Two branches (IB, LB) and three branches (IB, UB, LB) amplitude responses are obtained for the low and high Re values, respectively. The wake behind the cylinder with 2-DOF VIV undergoes the *mode-C* transition of 2D to 3D flow as opposed to the direct *mode-B* transition observed for transverse only VIV in the literature. The critical *Re* range of the 2D to 3D transition for the 2-DOF VIV cylinder at a reduced velocity of 6 is around 250, less than the 1-DOF VIV. Also, this range varies with the variation in $m^*$ and the streamwise to transverse oscillation frequency ratio ($f^*$). A $\lambda_z - \text{Re}$ map (where $\lambda_z$ represents the spanwise wavelength of the streamwise vortices) is proposed for the 2-DOF VIV, highlighting the different modes of transition obtained for combinations of $f^*$ and Re.

## I. INTRODUCTION

Vortex-Induced Vibration (VIV) can be defined as the response of the elastically mounted body to the incoming flow in terms of the structural oscillations (for rigid bodies) and/or deformations (elastic bodies). It has gained a significant amount of research interest due to its exciting flow physics and wide applications in many fields of engineering, such as marine engineering (riser tubes, offshore platforms, underwater structures), civil engineering (high-rise buildings, bridges), thermal engineering (heat-exchanger tubes, energy converters), electrical engineering (high-voltage towers, cables), etc. The practical applications of VIV have led to many experimental[1-7] and numerical studies[8-16] in the past. Williamson's study[17] on the stationary cylinder highlights the three-dimensionality behind the stationary cylinder when the Reynolds number (Re) based on the free-stream velocity and the cylinder diameter exceeds 178. However, experimental investigation of the flow behind the freely vibrating cylinder poses limitations due to the motion of the cylinder[18, 19]. People have used numerical simulations to investigate the three-dimensionality behind the freely oscillating cylinder[8-10, 20]. The mounting arrangement of the cylinder determines the degree-of-freedom (DOF) for the VIV system. When the cylinder is restricted to oscillate in the transverse direction only, it is known as the 1-DOF VIV system. A typical 1-DOF VIV amplitude response is divided into three-branch responses, i.e., the initial branch (IB), upper branch (UB), and the lower branch (LB).

Adding streamwise motion to the VIV (a 2-DOF VIV system) results in the suppression of a secondary vortex which influences the shedding phase and governs the sign of energy transfer[21]. In addition to the transverse oscillations, the cylinder exhibits periodic oscillations in the flow direction with a frequency two times the transverse oscillation frequency. The trajectory of the cylinder's oscillations under 2-DOF resembles figure "8". Jauvtis and Williamson[22, 23] designed a pendulum apparatus to achieve the 2-DOF motion to maintain the equal mass ratio and frequency ratio between the streamwise and the transverse directions. They reported that for the cylinder with a lower mass ratio ($m^* < 6$), a new response branch, "super-upper," exists, linked to a higher amplitude of oscillations in streamwise and transverse directions. In addition, if the mass ratio is greater than 6, the 2-DOF VIV response of the cylinder was identical to those of the cylinder with transverse only VIV. Thus, based on the study by Jauvtis and Williamson[23], the authors have selected the lower mass ratio ($m^* = 2.546$) for the present study. Further, Kang and Jia[24] demonstrated that for the 2-DOF VIV system, streamwise vibrations are in the form of multi-frequency vibrations. They found that under a particular combination of the frequency ratio and reduced velocities, the streamwise oscillation frequency spectrum contains the transverse frequency and the frequency twice. They further reported the "D," "egg," and "raindrop" shaped cylinder trajectories [significantly different from the traditional figure "8" and "new moon" shaped trajectories] which depended primarily on the selection of natural vibration frequency and reduced velocity. Dahl et al.[25] studied the effect of the streamwise and transverse natural frequency ratio on the system response and reported that fluid forces, structural responses, and wake patterns would vary significantly if the cylinder is permitted to move with combined streamwise and transverse oscillations.

Apart from the experimental studies, many researchers have employed computational fluid dynamics (CFD) tools to study the VIV of a circular cylinder using both two-dimensional and three-dimensional models. Leontini et al.[26] reported the branching behavior at low Reynolds number flows (Re = 200) for 1-DOF VIV systems. They obtained two response branches similar to the upper and lower branch (at high Reynolds number flows). They concluded that the three-dimensional flow branching behavior originates in the two-dimensional flow. Zhao et al.[20] presented the simulation results for the 1-DOF VIV and attempted to address the wake transition from two-dimensional to three-dimensional by varying the Reynolds number. They also captured the initial and lower branches (for Re = 150) and the upper branches (for Re = 1000) for the transverse oscillation amplitude response. Singh and Mittal[27] studied the hysteresis behavior in the oscillation amplitude response via numerical simulation of a 2-DOF VIV system at low Reynolds number flow. They reported the hysteresis in the lock-in range and obtained the P+S vortex shedding mode in the free vibrations for the first time. Lucor and Triantafyllou[28] studied the effect of varying the streamwise to transverse oscillation frequency ratio for a 2-DOF VIV system via two-dimensional simulations. They reported that the peak oscillation increases and shifts to the higher reduced velocities with the increase in the frequency ratio. Kondo[29] performed the 3D computations for the 2-DOF VIV system and captured the first and second excited vibrations of the streamwise direction. Navrose and Mittal[30] observed the hysteresis in the initial-upper branch transition and the intermittency in the upper-lower branch transition for a 2-DOF VIV system at Re = 1000. The numerical results by Gsell et al.[31] for the 2-DOF VIV system at Re = 3900 reported the variation in the phase difference between the streamwise and transverse motions across the lock-in range. When the streamwise frequency is two times the transverse frequency ($f^* = 2.0$), a perfect resonance occurs[32]. In the experimental and numerical investigations of 2-DOF VIV, Srinil et al.[33] obtained the eight-shaped oscillation trajectory for a wide range of reduced velocities indicating the dual resonance. Further, Bao et al.[34] also confirmed the existence of the dual-resonance over a wide range of $f^*$ for a 2-DOF VIV at Re = 150.

For a freely oscillating rigid cylinder, the transition to three-dimensionality influences the wake formation behind the cylinder, and the critical Reynolds number at which this transition to three-dimensionality takes place depends on the cylinder's oscillation amplitude. Roshko[35] reported the existence of the wake transition from two-dimensional to three-dimensional behind a stationary cylinder in the range of Re = 150-300. Later, Williamson[36] studied the distinct irregularities in the wake velocity signals, reported a discontinuous change in the wake formation at Re = 178, and named it mode-A instability. This mode-A instability exhibits the spanwise wavelength of about $\lambda_A = 3D \sim 4D$. On further increasing the Reynolds number (Re = 230-260), he reported the transition to mode-B instability with a spanwise wavelength of about $\lambda_A = 0.8D \sim 1D$. These modes have been confirmed by several other studies[17, 36-38]. Hover et al.[19] experimentally studied the three-dimensionality of mode transition in the wake of a flexibly-mounted rigid cylinder undergoing VIV. They observed the spanwise correlation loss near maximum amplitude conditions and highlighted the three-dimensional effects' sensitivity to the physical parameters. Further, Du et al.[10] and Zhao et al.[20] studied the modes of vortex formation and transition to three-dimensional wake via three-dimensional numerical simulations. They reported that the flow transition for VIV occurs via mode-B at around Re = 300, where the dominance of the streamwise vortices over the spanwise vortices becomes strong with an increase in the Reynolds number. Leontini *et al.*[8-9] performed the Floquet stability analysis of the three-dimensional transition in the wake of a cylinder forced to oscillate transversely. They showed that the amplitude of cylinder oscillation causes the critical Reynolds number for mode A to increase above the critical Reynolds number for mode B. Hence, the transition in the oscillating cylinders occurs via mode-B.

In the case of long, slender ocean structures like risers, mooring lines, and cables, vortex-induced forces can excite many natural frequencies, frequently at high mode numbers. Thus, their vibratory response can differ from short, flexibly placed cylinders, including the creation of traveling waves and multiple frequency excitation, especially in sheared currents. However, long structures' response exhibit similarities in several ways during the experimental lab studies. Vandiver *et al.*[39, 40] observed cylinder motions and third-harmonic force components on a long, flexible pipe. Lucor *et al.*[41], Vandiver *et al.*[39], and Mukundan *et al.*[42] found that figure-eight shapes, seen in the two-degree-of-freedom response of a rigid cylinder, are also found in the traveling waves along a long, flexible pipe, accompanied by significant third harmonics. Bao *et al.*[43], Thapa *et al.*[44], and Zhang *et al.*[45] reported the modes of the wake three-dimensionality similar to the elastically mounted rigid cylinder at low Reynolds numbers. Although, the current study primarily focuses on the mode dynamics of the elastically mounted rigid cylinder.

Previous research has shown that the streamwise motion considerably impacts the motion trajectory and hydrodynamic forces on the cylinder. Notably, there have been several studies on the 2-DOF VIV; nevertheless, the majority examined the VIV oscillation amplitude-frequency response in conjunction with vortex shedding modes. To the best of the authors' knowledge, no systematic study has been undertaken on the correlation (mapping) between the wake transition modes and the Reynolds number for the 2-DOF VIV system. In this paper, we address the question of the relationship between the 2-D to 3-D transition and the critical Reynolds number at which this transition occurs for the 2-DOF VIV system using high-fidelity numerical simulations. Further, the aim is to shed light on the effects of the different VIV parameters associated, such as frequency ratio, mass ratio, and damping ratio, on the onset of the three-dimensionality in the wake flow behind the oscillating cylinder under the 2-DOF system. In particular, the wake flow behind the cylinder undergoing the 2-DOF motion may transit to 3D via different modes of three-dimensionality (i.e., *mode-A, B, and C*), depending on the combination of the Reynolds number and the streamwise to transverse oscillation frequency ratio. Finally, the present study summarizes this dependence of mode transition via the study of the spanwise wavelength ($\lambda_z$) of the streamwise vortices by proposing a $\lambda_z - \text{Re}$ map similar to the one obtained by Williamson[17] for the wake behind a stationary circular cylinder, which has not been reported earlier.

## II. PROBLEM STATEMENT

The present work examines the VIV characteristics of an elastically mounted circular cylinder at Reynolds numbers (based on the cylinder diameter, D) ranging from 150 to 1000. The equations are solved numerically over a three-dimensional computational domain, shown in Fig. 1(a). The cylinder is elastically mounted by the spring-damper arrangement to vibrate in the streamwise and transverse direction only. The springs in both directions are assumed to be linear and exhibit the same stiffness. The circular cylinder has a non-dimensional mass ratio $(m^*)$ of 2.546, a length-to-diameter ratio, $L/D = 9.6$ and the structural damping ratio $(\zeta = 0)$. The computational domain spans between $-20 \leq x/D \leq 30$, $-20 \leq y/D \leq 20$, and $0 \leq z/D \leq 9.6$, considering the center of the cylinder located at the origin. The main objective of the study is to numerically study the 3D flow characteristics associated with a 2-DOF VIV cylinder within the subcritical Re regime. Besides the study of 3D flow characteristics, there is also a particular focus on essential aspects of the flow, such as the transition from 2D to 3D, the correlation between the forces and the oscillation amplitude, and the variation of the pressure distribution over the cylinder surface. The incoming flow is assumed to be steady and uniform, while the outlet is treated with the advective boundary condition for the flow velocity. The surface of the cylinder is assumed to be smooth and treated with no-slip boundary conditions, i.e., the fluid velocity along the surface of the cylinder is the same as the cylinder's vibration speed. The free-slip boundary condition is applied on the two spanwise and the transverse boundaries, popularly used[20, 30].

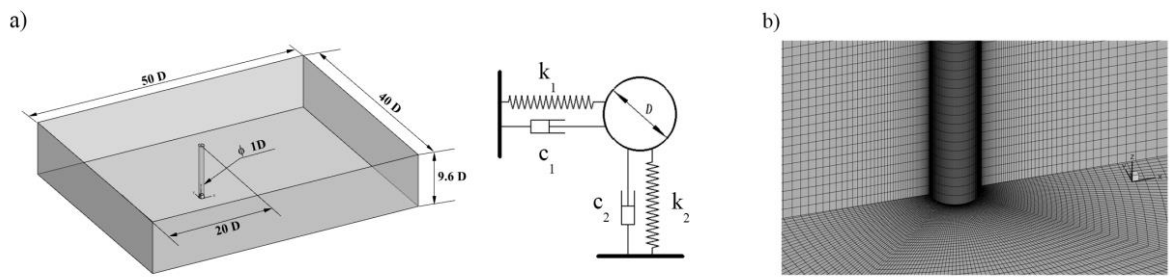

**FIG. 1.** (a) Schematic of the computational domain, (b) computational mesh used for VIV of a circular cylinder

## III. NUMERICAL DETAILS

### A. Flow Solver

The simulation of the flow is governed by the unsteady 3D incompressible Navier-Stokes (NS) equations written as

$$\nabla \cdot \vec{v} = 0 \qquad (1)$$

$$\rho_f \left[ \frac{\partial \vec{v}}{\partial t} + (\vec{v} \cdot \nabla)\vec{v} \right] = -\nabla p + \mu_f \nabla^2 \vec{v} \qquad (2)$$

where $\vec{v}$ is the velocity vector, $\rho_f$ is the fluid density, $p$ is the static pressure, and $\mu_f$ is the dynamic viscosity of the fluid. The Reynolds number is defined based on the cylinder diameter and the kinematic viscosity ($\nu$) as $\text{Re} = \frac{UD}{\nu}$. Open source CFD solver, OpenFOAM[46], is used to solve the flow equations. The pressure-velocity coupling is felicitated using the PIMPLE algorithm. Second-order discretization schemes are invoked to address all the spatial and temporal terms in the governing equations.

### B. Structural Solver

The motion of a rigid cylinder mounted elastically on the spring-damper arrangement can be represented by the following mathematical model,

$$m\ddot{x}_i + c_i \dot{x}_i + k_i x_i = F_i \qquad (3)$$

where $x_1 = x$ and $x_2 = y$ represent the cylinder displacements in the x and y-directions, respectively. $c_i$, $k_i$, and $F_i$ are the damping coefficient, structural stiffness, and the force component (pressure + viscous forces) in the $x_i$-direction, respectively. The springs are assumed to be linear and have the same stiffness in streamwise and transverse directions, i.e., $k_1 = k_2 = k$. This also implies that the structure's natural frequency is similar in both directions, i.e., $f_{nx} = f_{ny} = f_n$. Following the previous literature, the motion of such a system can be represented using some of the non-dimensional parameters such as non-dimensional mass-ratio $\left( m^* = \frac{m}{\frac{\pi}{4}\rho D^2 L} \right)$, non-dimensional spring constant $\left( k^* = \frac{k}{\rho U_\infty^2 L} \right)$, non-dimensional damping coefficient $\left( C^* = \frac{c}{\rho U_\infty D L} \right)$, non-dimensional reduced velocity $\left(U_r = \frac{U_\infty}{f_n D}\right)$, non-dimensional frequency $\left(f = \frac{f_s D}{U}\right)$, frequency-ratio $\left(f^* = \frac{f_s}{f_n}\right)$, non-dimensional time $\left(\tau = \frac{tU}{D}\right)$, non-dimensional mean streamwise oscillation amplitude $\left(A_x^* = \frac{(X/D)_{max} + abs(X/D)_{min}}{2}\right)$, and non-dimensional mean transverse oscillation amplitude $\left(A_y^* = \frac{(Y/D)_{max} + abs(Y/D)_{min}}{2}\right)$.

## C. Fluid-Structure Coupling and Mesh Motion

Eqn. (3) provides the updated position of the cylinder based on the forces acting on it. Thus, it requires the forces information simultaneously from the fluid solver, leading to the need for efficient coupling between the fluid and structural equations. The fluid force is calculated at each time step by solving Eqn. (1) and (2), and then the motion of the cylinder is determined by solving the Eqn. (3) using the fourth-order Runge-Kutta scheme. We have utilized the weakly coupled form of the solvers as followed by the other researchers[47-49]. The time-step size is kept sufficiently small for all the cases for better convergence for the coupling. The mesh motion is calculated using Laplace's equation,

$$\nabla \cdot (\gamma_m \nabla z) = 0 \qquad (4)$$

Where, $\gamma_m$ is the diffusion coefficient, and $z$ is the mesh-cell center displacement field. The mesh motion is distributed using the inverse mesh diffusion model based on the inverse distance from the cylinder[50-51]. The diffusivity field is calculated using the quadratic relation on the inverse of the cell center distance ($l$) to the nearest boundary, $1/l^2$.

## D. Grid-Independence Study and Error Analysis

To examine the independence of the calculated findings from the grid size, we have conducted the grid-independence investigation at Re = 150 and Re = 1000 to cover the entire Reynolds number range. The mass ratio is 2, the damping ratio is zero, and the frequency ratio is one. The cylinder length is 9.6 times the cylinder's diameter (D). Mesh is constructed using ANSYS-ICEM CFD[52] by invoking O-grid blocking in the near cylinder region, as depicted in Figure 1. (a). With a cell expansion ratio of 1.02 in the small square region surrounding the cylinder, the height of the initial layer at the cylinder's surface is kept low. To reduce computing cost, the remainder of the computational domain is filled with hexahedral cells with an expansion ratio of 1.2.

Three different grid sizes, i.e., Grid-1 [with 882,080 cells], Grid-2 [with 1,734,200], Grid-3 [with 3,411,828], and Grid-4 [with 6,463,859] are simulated and the r.m.s. value of the lift coefficient ($C_{L_{rms}}$) and the oscillation amplitudes are compared to assess the grid independence in Table I at Re = 150. Figure 2 shows the results of the grid-independence study via the time history of the transverse oscillation and the corresponding lift coefficient and its variation with the mesh sizes for Re = 150 and 1000, respectively. Also, Tables I and II tabulate the results obtained for r.m.s. of lift coefficient and the mean oscillation amplitude for the different grid sizes. We observe that the difference between Grid-3 and Grid-4 drops to less than 2% for both Reynolds numbers. Hence, Grid-3 is selected to perform this study.

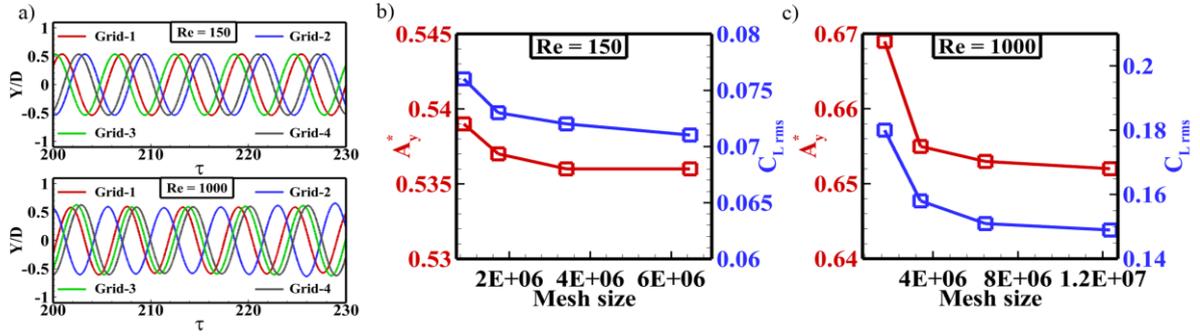

**FIG. 2.** Grid Independence study for the moving cylinder at Re = 150 and 1000: (a) Time-history of the transverse displacement; and Variation of mean transverse oscillation amplitude and the rms value of corresponding lift coefficient at: (b) Re = 150, c) Re = 1000. [ $m^* = 2, \zeta = 0, f^* = 1.0, U_r = 6, L/D = 9.6$ ]

**Table I.** Grid Independence study for the moving cylinder at Re = 150 [Values in bold show the selected grid]

| Grid | No. of cells | $C_{L\,rms}$ | % Change in $C_{L\,rms}$ value | $A_y^*$ | % Change in $A_y^*$ value |
|---|---|---|---|---|---|
| Grid-1 | 882,080 | 0.076 | - | 0.539 | - |
| Grid-2 | 1,734,200 | 0.073 | 3.356 | 0.537 | 0.271 |
| **Grid-3** | **3,411,828** | **0.072** | **1.988** | **0.536** | **0.151** |
| Grid-4 | 6,463,859 | 0.071 | 1.364 | 0.536 | 0.048 |

**Table II.** Grid Independence study for the moving cylinder at Re = 1000 [Values in bold show the selected grid]

| Grid | No. of cells | $C_{L\,rms}$ | % Change in $C_{L\,rms}$ value | $A_y^*$ | % Change in $A_y^*$ value |
|---|---|---|---|---|---|
| Grid-1 | 1,734,200 | 0.180 | - | 0.669 | - |
| Grid-2 | 3,411,828 | 0.158 | 12.170 | 0.655 | 1.994 |
| **Grid-3** | **6,463,859** | **0.151** | **4.143** | **0.653** | **0.294** |
| Grid-4 | 12,330,600 | 0.149 | 1.677 | 0.652 | 0.258 |

Further, to gain more confidence in the numerical results, we have also calculated the Grid Convergence Index (GCI) proposed by Roache[53-54], and extensively used in the past literature. Appendix A.1 shows a detailed description of the GCI calculations. The GCI calculations are performed over the r.m.s. value of the lift coefficient ($C_{L_{rms}}$) for Re = 1000. The factor of safety value is taken as 1.25[55]. The results for the GCI calculations are tabulated in Table III. These calculations confirm that the grid is nicely resolved.

**TABLE III.** Richardson error estimation and grid-convergence index for three sets of grids

|  | $r_{CB}$ | $r_{DC}$ | o | $\varepsilon_{CB}$ | $\varepsilon_{DC}$ | $E_2^{coarse}$ | $E_1^{fine}$ | $GCI^{coarse}$ | $GCI^{fine}$ |
|---|---|---|---|---|---|---|---|---|---|
| $C_{L\,rms}$(Re1000) | 1.25 | 1.25 | 1.916 | -0.044 | -0.013 | -0.126 | -0.024 | 15.750 % | 3.100 % |

## E. Numerical Validation

To check the accuracy of the computed results, we have validated our numerical setup with the available published literature[30, 34]. A rigid cylinder is mounted on the elastic supports to vibrate in streamwise and transverse directions. As the study deals with the range of Reynolds numbers, the authors have validated the numerical setup for the two extremes of the range, i.e. Re = 150 (with a comparison of Bao et al.[34]) and Re = 1000 (with a comparison of Navrose and Mittal[30]). The mass ratio of the cylinder is 2, with a frequency and damping ratio of 1.0 and 0, respectively. Figure 3 depicts the two-different VIV response branches for the Re = 150 compared with the published data by Bao et al.[34]: the initial branch (IB, $U_r \leq 4$), the lower branch (LB, $4 \leq U_r \leq 8$), along with the desynchronized branch (DS, $U_r \geq 8$). The results obtained from the numerical simulations agree well with the published literature for all three branches. To check for the lock-in regime, Fig. 3(c) reports the frequency response of the cylinder over the different reduced velocities. The lock-in can be identified with the frequency vs. reduced velocity curve being flat around the natural frequency.

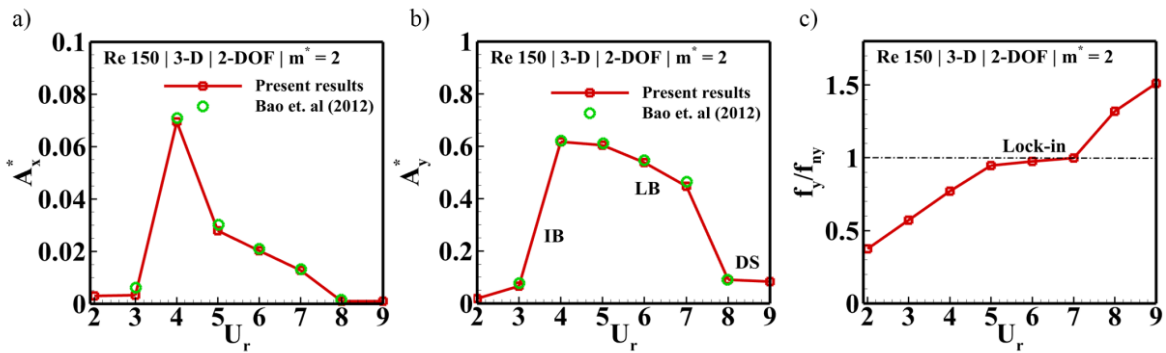

**FIG. 3.** Numerical validation with the available literature at a low Reynolds number of Re = 150: (a) Streamwise mean oscillation amplitude ($A_X^*$), (b) transverse mean oscillation amplitude ($A_Y^*$), and (c) the transverse frequency response over the range of reduced velocities. [$m^* = 2, \zeta = 0, f^* = 1.0, \text{Re} = 150, L/D = 9.6$]

Further, to confirm the accuracy of the simulations at the higher Reynolds number, we have also validated our results with the results obtained by Navrose and Mittal[30]. The cylinder's mass ratio is 10, and the damping ratio is 0, with a frequency ratio of 1.0. The cylinder spans four times the diameter in the spanwise direction. Figure 4 compares the amplitude responses (both streamwise and transverse amplitude responses) and the r.m.s. values of the corresponding aerodynamic lift and drag coefficients for the range of reduced velocities. The results agree well with the literature; the slight difference can be attributed to differences in the numerical schemes and the initial conditions.

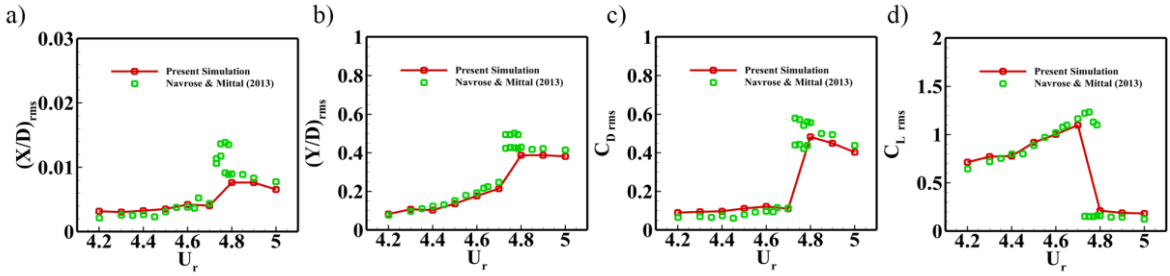

**FIG. 4.** Validation study at different reduced velocity for Re = 1000: (a) r.m.s. value of streamwise oscillation amplitude, (b) r.m.s. value of the transverse oscillation amplitude, (c) r.m.s. value of the surface averaged drag coefficient ($C_{D_{rms}}$), and (d) r.m.s. value of the surface averaged lift coefficient ($C_{L_{rms}}$). [$m^* = 10, \zeta = 0, f^* = 1.0, \text{Re} = 1000, L/D = 4.0$]

## IV. RESULTS AND DISCUSSION

In this section, we comprehensively investigate the flow characteristics and associated VIV responses of an elastically mounted circular cylinder. The cylinder is mounted using a spring and damper system. It is free to oscillate in transverse and inline directions under the aerodynamic forces, making it a 2-DOF VIV system. The article has two parts: the first part (sections IV.A and IV.B) presents the VIV response of the cylinder at a low Re of 150 and a higher Re of 1000 in detail via means of two-dimensional and three-dimensional simulations. This section highlights the various flow characteristics associated with the oscillating cylinder and checks the validity of the two-dimensional simulations for these two Re values. The latter part (section IV.C) reports a detailed three-dimensional analysis to assess the transition of the wake-flow from a fully two-dimensional wake to a fully three-dimensional wake over the range of Re (i.e., Re = 150-1000). The effect of the various VIV parameters, e.g. mass-ratios ($m^*$), frequency-ratios ($f^*$), and damping ratios ($\zeta$) on the three-dimensionality of the cylinder wake, is also duly reported in the subsections of section IV.C.

## A. VIV at a low Reynolds number of 150

### *A.1. Oscillation Amplitude and Frequency Response*

Figure 5 (a) and (b) demonstrate the variation of the mean streamwise ($A_x^*$) and transverse ($A_y^*$) oscillation amplitude over the range of reduced velocities ($2 \leq U_r \leq 12$) at a Reynolds number of 150 for the 3-D model, and the results are compared with the 2D numerical results. The frequency ratio ($f^*$) is kept at 1. It can be seen that the 3D numerical results are identical to the 2D numerical results over the complete range of $U_r$ this Reynolds number. Further, the shapes of the oscillation-amplitude response curves are compared with those obtained by Bao et al.[34] from 2D CFD simulations at Re = 150 for $m^* = 2$ and found to be similar. Bao et al.[34] reported that the higher value of streamwise oscillation amplitude can be attributed to the lower mass ratio value. Figure 7 (c) and (d) depict the streamwise frequency ($f_{nx}$) and the transverse-frequency ($f_{ny}$) response normalized with the natural frequency ($f_n$) of the system. The streamwise frequency is almost twice the transverse frequency, confirming the dual resonance occurrence widely reported for 2-DOF VIV systems[32-34]. From the amplitude response and frequency response, one can identify the lock-in region by the high oscillation amplitudes and synchronization between the response frequency and the system's natural frequency, and the lock-in spans over $5 \leq U_r \leq 7$. The response frequency is almost constant in the lock-in region (twice the natural frequency for the streamwise oscillations and almost equal to the natural frequency for the transverse oscillations) and varies linearly with the reduced velocity outside the lock-in regime.

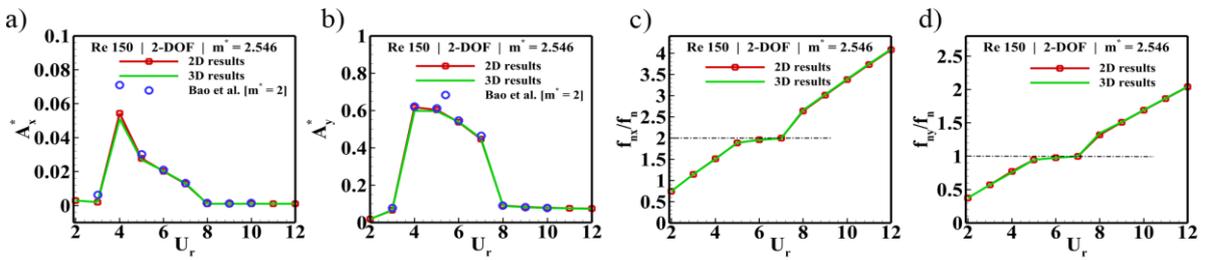

**FIG. 5.** Amplitude and frequency responses over the range of reduced velocities for the 2-DOF VIV cylinder at Re = 150. (a) streamwise oscillation amplitude, (b) transverse oscillation amplitude, (c) streamwise oscillation frequency response, and (d) transverse oscillation frequency response [$m^* = 2.546$, $\zeta = 0$, Re = 150]

Figure 6 presents the time histories of the total lift coefficient ($C_L$) with the corresponding transverse oscillation amplitude and their frequency spectrum for different reduced velocities (Fig. 7). The oscillation amplitude is lower for lower reduced velocities ($U_r < 4$) due to the low lift coefficient values on the cylinder surface. At the reduced velocity of 4, the cylinder starts to oscillate with a larger oscillation amplitude due to the high value of the lift coefficient. As the reduced velocity increases ( $5 \leq U_r \leq 7$ ), the oscillation amplitude remains almost constant, even with the continuous decrease in the lift coefficient. This can be attributed to the synchronization of the system's frequency response with its natural frequency, as seen in Fig. 5 (c) and (d). The lift force on the cylinder remains in phase with the transverse oscillation up to this range of the reduced velocities. As reduced velocity reaches 8, the cylinder oscillates out-of-phase with the lift force. For the reduced velocities $U_r \geq 8$, the oscillation amplitude and the lift coefficient reduce to a lower value and remain constant for the remaining reduced velocities.

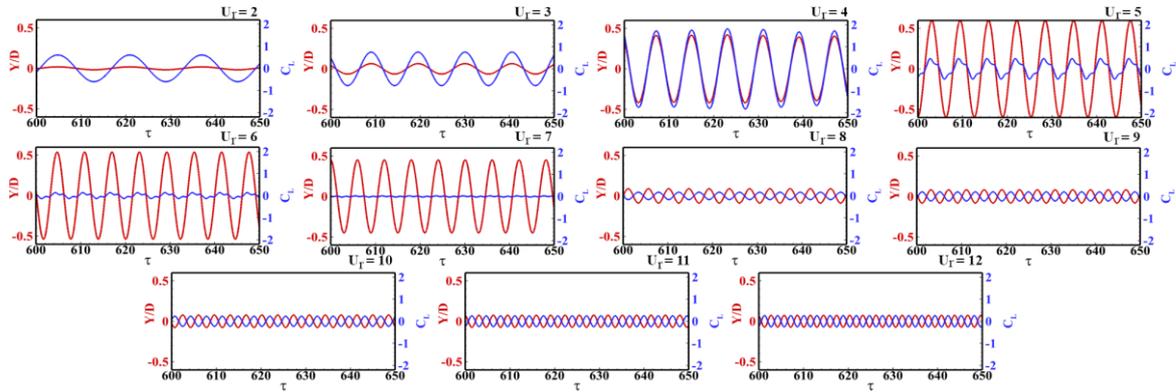

**FIG. 6.** Time histories of the transverse oscillation amplitude and the corresponding total lift coefficient over the 2-DOF VIV system at Re = 150. [ $m^* = 2.546$, $\zeta = 0$, Re = 150]

Figure 7 shows the frequency response of the oscillating cylinder for its streamwise oscillation amplitude ( $f_{x/D}$ ), transverse oscillation amplitude ( $f_{y/D}$ ), and the total lift coefficient ( $f_{Cl}$ ) at different reduced velocities. The cylinder oscillates in the transverse direction with the same frequency as the frequency of the lift coefficient throughout the reduced velocities. The lift coefficient frequency spectrum suggests that the single peak frequency dominates its temporal evolution for lower values of the reduced velocities ( $U_r < 4$ ). As the reduced velocity increases ( $5 \leq U_r \leq 7$ ), another peak frequency (third harmonics of the lift coefficient) dominates during the lock-in regime, and the cylinder oscillates out-of-phase with the lift force. Jauvtis & Williamson[23] and Dahl et al.[25, 32, 56-57] for the cylinder's combined in-line and crossflow motions also observed this significant third harmonic component of lift. Dahl *et al.*[56] showed that when the cylinder sweeps a figure-eight motion trajectory and moves upstream at the two extreme ends of the orbit (counter-clockwise motion), additional vorticity is generated, resulting in multiple vortex formation and high force harmonics. Dahl *et al.*[57] used a simple potential flow construction to explain how the vortex formation pattern, observed for the associated counter-clockwise orbital motion at $U_r$ = 6.4, cancels lift and drag forces in phase with acceleration, resulting in an effective added mass close to zero in both the in-line and crossflow directions. When structural damping is low, the crossflow fluid force in phase with velocity must be close to zero for free, steady-state oscillations; thus, the whole first harmonic of the crossflow force must be near zero. This is why the third harmonic lift force dominates in this regime. In the post-lock-in regime, the dominance of this higher frequency reduces, and the cylinder again oscillates with the single dominant peak frequency.

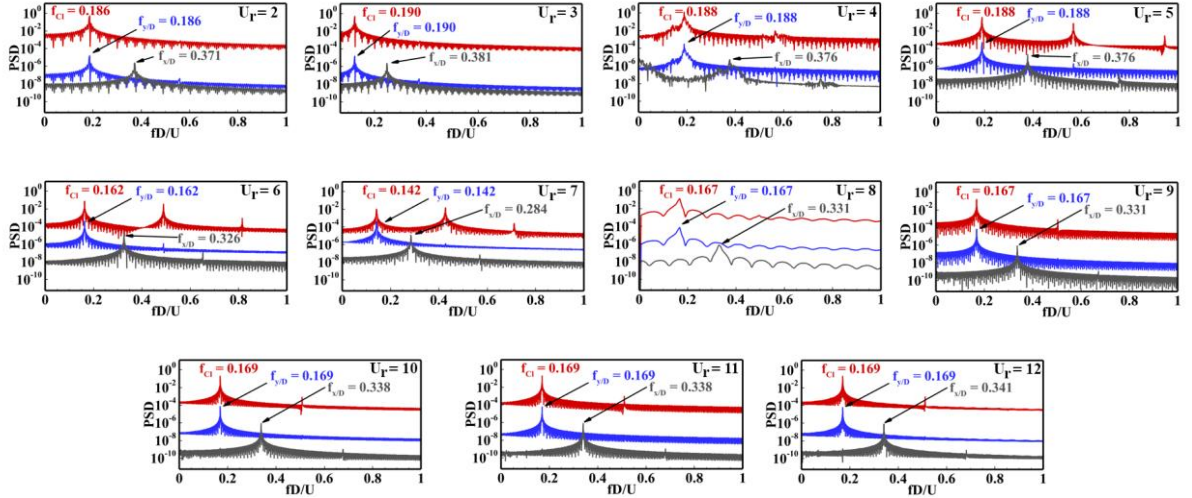

**FIG. 7.** Frequency spectrum of the lift coefficient (in red), transverse oscillation amplitude (in blue), and the streamwise oscillation (in grey) for the 2-DOF VIV system at Re = 150 for different reduced velocities. [ $m^* = 2.546$, $\zeta = 0$, Re = 150]

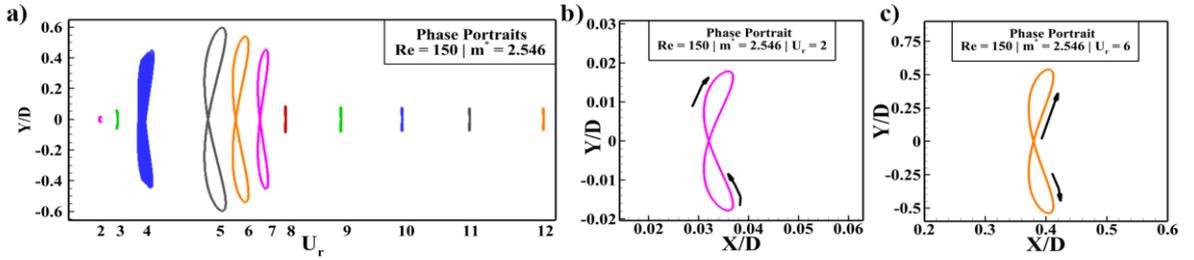

**FIG. 8.** (a) Variation of the orbital trajectories with the reduced velocity, (b) clockwise figure-eight trajectory for $U_r = 2$, and (c) counter-clockwise figure-eight orbital path for $U_r = 6$ [ $m^* = 2.546$, $\zeta = 0$, Re = 150]

The past literature[32, 58-61] reported that the orbital trajectory is critical in determining the energy transfer between the fluid and the structure. The variation of the orbital trajectories with reduced velocity is given in Fig. 8. As observed, the cylinder sweeps figure-eight-shaped trajectories for almost all the reduced velocities, which reconfirms the occurrence of the dual resonance ( $f_{nx} / f_{ny} = 2$ ). For the lower reduced velocities ( $U_r \leq 4$ ), the cylinder sweeps a clockwise trajectory, which indicates damping offered by the flow to the cylinder vibrations[48]. The cylinder follows a counter-clockwise orbit for the remaining higher values of the velocities ( $4 < U_r \leq 12$ ), indicating the energy transfer from the fluid to the cylinder[32]. The counter-clockwise trajectory during the lock-in regime agrees well with the available literature[59].

## A.2. Flow Characteristics

This sub-section reports the 3D and 2D results of the 2-DOF VIV system at Re=150. The mass ratio ($m^*$) is taken at a low value of 2.546, and the damping coefficient is taken at 0 to maximize the VIV oscillations. Figure 9 shows the flow distribution employing instantaneous z-vorticity, time-averaged velocity streamlines, and the corresponding Reynolds shear stress ($<u'v'/U_\infty^2>$) on the 2D slice extracted at the midspan of the cylinder (z/D = 4.8) for different reduced velocities. The vorticity contours confirm the 2S shedding corresponding to the shedding of one pair of vortices in one period of oscillation. The mean velocity streamlines show a pair of primary vortices downstream the cylinder for $U_r = 2, 7,$ and 10. At the same time, it disappears for $U_r = 5$ due to high oscillations during lock-in. Authors firmly believe that the disappearance of the pair of the primary vortices downstream of the cylinder can also be used to visualize the lock-in phenomena for the 1-DOF and 2-DOF VIV.

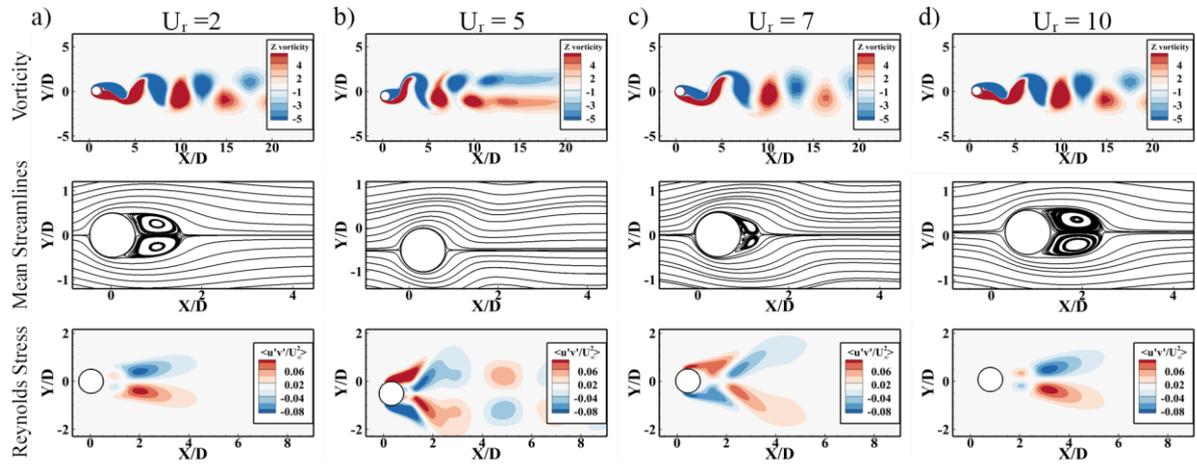

**FIG. 9.** Instantaneous z-vorticity contours, mean streamlines drawn from the time-averaged velocity field, and normalized Reynold's shear stress for 2-DOF VIV of a circular cylinder at Re = 150 for: (a) $U_r = 2$, (b) $U_r = 5$, (c) $U_r = 7$, and (d) $U_r = 10$ [$m^* = 2.546$, $\zeta = 0$, 3D]

Figure 9 depicts the anti-symmetric distribution of the Reynolds shear stress about the cylinder-wake axis for all the reduced velocities. Reynolds shear stress ($<u'v'/U_\infty^2>$) is calculated by taking the mean of the product of the fluctuating components of the axial and transverse velocity (u' and v'), normalized with the square of the freestream velocity ($U_\infty$). The statistics are collected after the non-dimensional time, $\tau = 300$, when the flow characteristics show the quasi-steady periodic behavior. The lower values of Reynolds stress near the cylinder (as for the $U_r = 2$ and 10) represent the weakening of the vortex shedding and supporting the stabilized wake. For $U_r = 5$, the higher values of Reynolds stress are found over the cylinder, denoting the high disturbance in the cylinder wake due to high-amplitude cylinder oscillations.

Further, to represent the 3-D vortex patterns in the cylinder wake, we have used the method proposed by Jeong and Hussain[62], in which the iso-surfaces of the second eigenvalue ($\lambda_2$) of the tensor $S^2 + \Omega^2$, where $S$ and $\Omega$ are the symmetric and antisymmetric parts of the velocity gradient tensor $\nabla u$, is used to identify the vortex core. Figure 10 displays the non-dimensional eigenvalue $e_2 = -\lambda_2/(U/D)^2$ for four different reduced velocities, corresponding to three different VIV branches ($U_r = 2$ corresponding to the initial branch; $U_r = 5$ and 7 corresponding to the lower branch; and $U_r = 10$ corresponding to the desynchronized branch of VIV). We can observe that the flow in the wake is two-dimensional at this Reynolds number (Re = 150) as the vortices are parallel to the cylinder span throughout the whole range of the reduced velocities. Also, the cylinder sheds one pair of vortices in one period of vibration and shows a 2S shedding pattern.

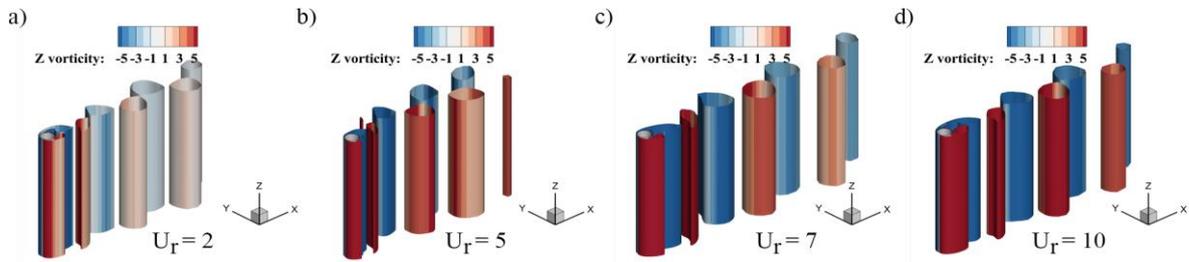

**FIG. 10.** Iso-surface of eigenvalue $e_2 = 0.01$ for VIV of a cylinder at Re = 150: (a) $U_r = 2$, (b) $U_r = 5$, (c) $U_r = 7$, and (d) $U_r = 10$ [$m^* = 2.546$, $\zeta = 0$]

## B. VIV at Reynolds number of 1000

### B.1. Oscillation Amplitude and Frequency Response

Figure 11 reports the variation of the mean transverse oscillation amplitude and the frequency response with the reduced velocities. The response is compared with the two-dimensional simulation results and the experimental and numerical results obtained by the other researchers for the 1-DOF system[2, 20], and the 2-DOF system[23]. In the experimental results[2, 23], the Reynolds number is varied from 2,000 to 12,000 while it remains constant (at Re = 1000) in the present numerical study. Khalak and Williamson[3] in their experimental study of the transversely oscillating cylinder, defined the amplitude response of a 1-DOF VIV system as three branch response: initial branch ($U_r < 4$, IB), where the oscillation amplitude increases with the increase in the reduced velocity, upper branch ($4 \leq U_r \leq 6$, UB) where the oscillation amplitude is highest and above 0.8 times of the cylinder diameter, and the lower branch ($6 < U_r \leq 10$, LB) where the oscillation amplitude is independent of the reduced velocity. The transverse oscillation amplitude for the 2-DOF VIV system obtained in the present simulations is similar to that observed in the experimental study for a transversely oscillating cylinder (1-DOF VIV system).

Jauvtis and Williamson[23] studied the 2-DOF VIV for a circular cylinder with a mass ratio of 2.6. They reported the existence of the 'supper-upper' branch related to the maximum streamwise and transverse oscillation amplitude response. This branch is not obtained in our numerical simulations because the experimental study was conducted by varying the Reynolds number to achieve the different reduced velocities. In contrast, in the numerical study, we have kept the Reynolds number fixed at Re = 1000. The response of 2-DOF VIV with a low mass-ratio cylinder at a fixed Reynolds number resembles the 1-DOF VIV response where the cylinder is allowed to oscillate only in the transverse direction.

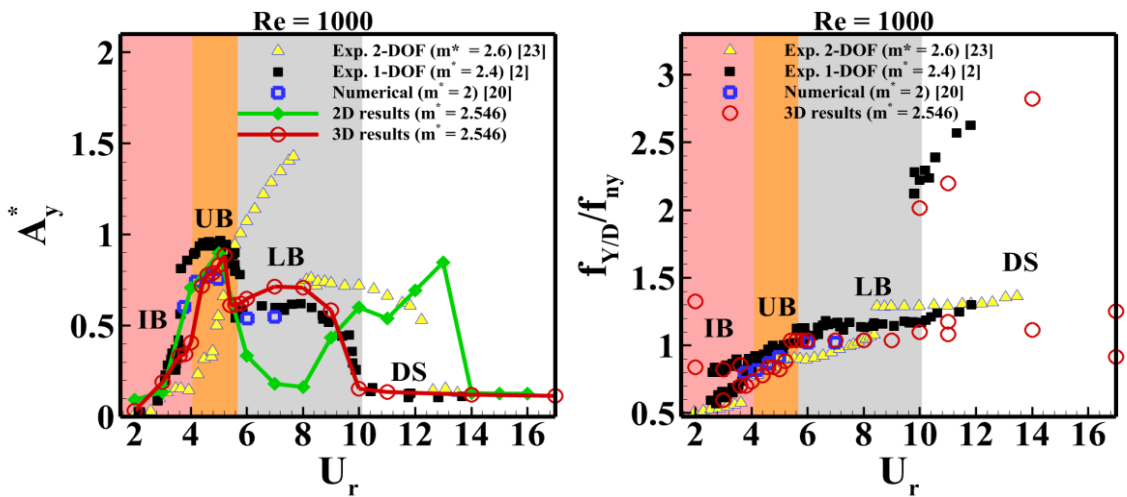

**FIG. 11.** Transverse oscillation amplitude and frequency response of a cylinder undergoing 2-DOF VIV. [ $m^* = 2.546, \zeta = 0, f^* = 1.0, \text{Re} = 1000$ ]

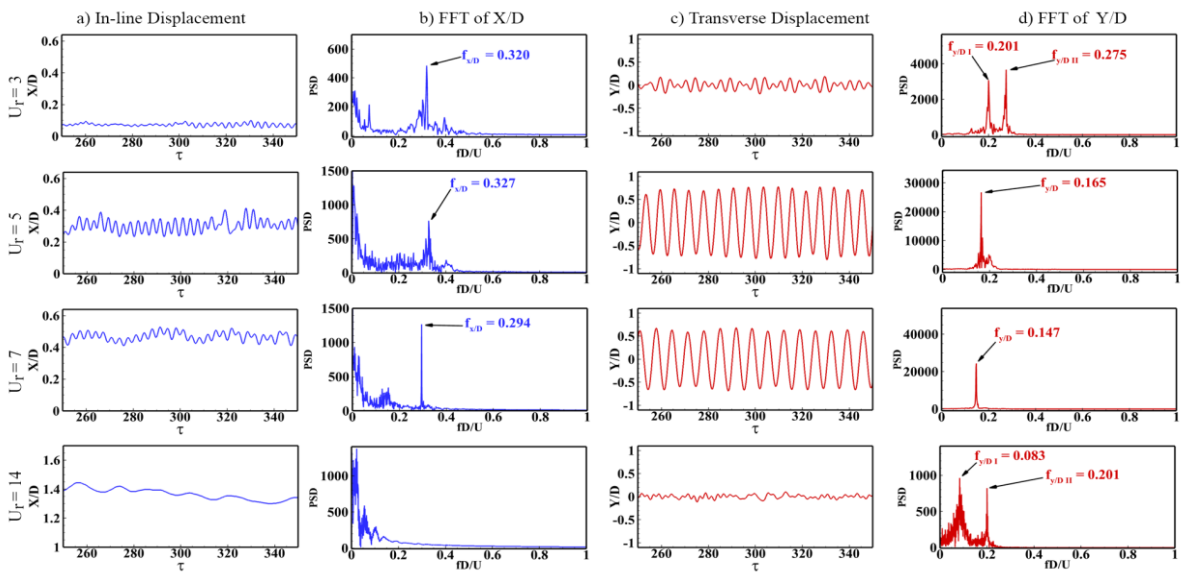

**FIG. 12.** Time-history of the streamwise and the transverse oscillation amplitude and their corresponding frequency spectrum: (a) In-line displacement, (b) FFT of X/D, (c) Transverse displacement, (d) FFT of Y/D. [ $m^* = 2.546, \zeta = 0, f^* = 1.0, \text{Re} = 1000$ ]

Further, Fig. 11 also compares the 2D numerical results with the 3D numerical results for the transverse oscillation amplitude of the cylinder under 2-DOF VIV. The 2D results overpredict the oscillation amplitude for the lower reduced velocities while heavily underpredicting the lower branch due to dominant three-dimensionality in the wake. The frequency response in Fig. 11 shows that the cylinder oscillations are governed by the number of frequencies during the initial branch, while the upper and lower branch response is governed by the single peak frequency, which is almost close to the natural frequency of the system. At the end of the lower branch, other high-frequency peaks appear and dominate the cylinder's oscillations.

Figure 12 provides the time histories of the oscillation amplitude in the streamwise and transverse directions with their corresponding frequency spectrum for some typically reduced velocities corresponding to different amplitude response branches. This also supports our preceding observations from the frequency response, i.e. the flow in the IB ($U_r = 3$) and DS ($U_r = 14$) is governed by the number of dominant frequencies, and the oscillation amplitude loses the periodicity. While a single harmonic peak governs the transverse oscillation of the cylinder for the UB ($U_r = 5$) and LB ($U_r = 7$), the oscillation amplitude response is periodic. Also, during the UB and LB ($U_r = 7$), the cylinder oscillates with the streamwise oscillation frequency, which is twice that of the transverse frequency. It also confirms the dual-resonance nature of these branch responses.

## B.2. Flow Characteristics

This sub-section presents the 2-DOF VIV results at Re = 1000 when the wake flow is assumed to be entirely three-dimensional[17] for reduced velocities ranging from 2 to 17 with an interval of 1. Figure 13 shows the flow visualization behind the cylinder oscillating under 2-DOF VIV conditions via instantaneous vorticity, time-averaged velocity streamlines, and the Reynolds stress at the z-plane mid-span. For the lower reduced velocities, the vortex shedding is weak, and the vortices form the 2-S type of shedding. As the reduced velocity increases, the vortex shedding becomes chaotic (as can be seen in the instantaneous spanwise vorticity contours behind the cylinder in Fig. 13), and the wake is filled with the dominant streamwise vortices (can be identified by looking at the dominant axial rib shaped vortical structures via the iso-surfaces of the non-dimensional eigen value ($e_2$) in Fig. 14 which represent the streamwise vortical structures) and the weak spanwise vortices (due to the associated vortex dislocation in the spanwise vortices).

The time-averaged velocity streamlines behind the cylinder denote the presence of the pairs of the primary vortices near the rear side of the cylinder, which represents the stabilized wake associated with the lower oscillation amplitudes. These pairs disappear when the cylinder oscillates with the higher oscillation amplitudes, typically during the lock-in in the upper and lower branch. To further access the disturbance behind the cylinder wake, Reynolds shear stress is plotted in Fig. 13. The distribution of the Reynolds stress is anti-symmetric about the wake axis. The peak values of the Reynolds stress component are near the cylinder surface for the upper and lower branches, representing the maximum disturbance created by the cylinder movement in these branches. Figure 14 depicts the iso-surface of the non-dimensional eigenvalue ($e_2$) colored with the spanwise vorticity at Re = 1000. It confirms that the wake behind the cylinder at this Reynolds number is entirely three-dimensional, and the spanwise dislocations of vortex tubes are dominant for the upper and lower branch. Also, one can observe a wider wake behind the cylinder for the upper and lower branch than the other two branches (refer to Fig. 13).

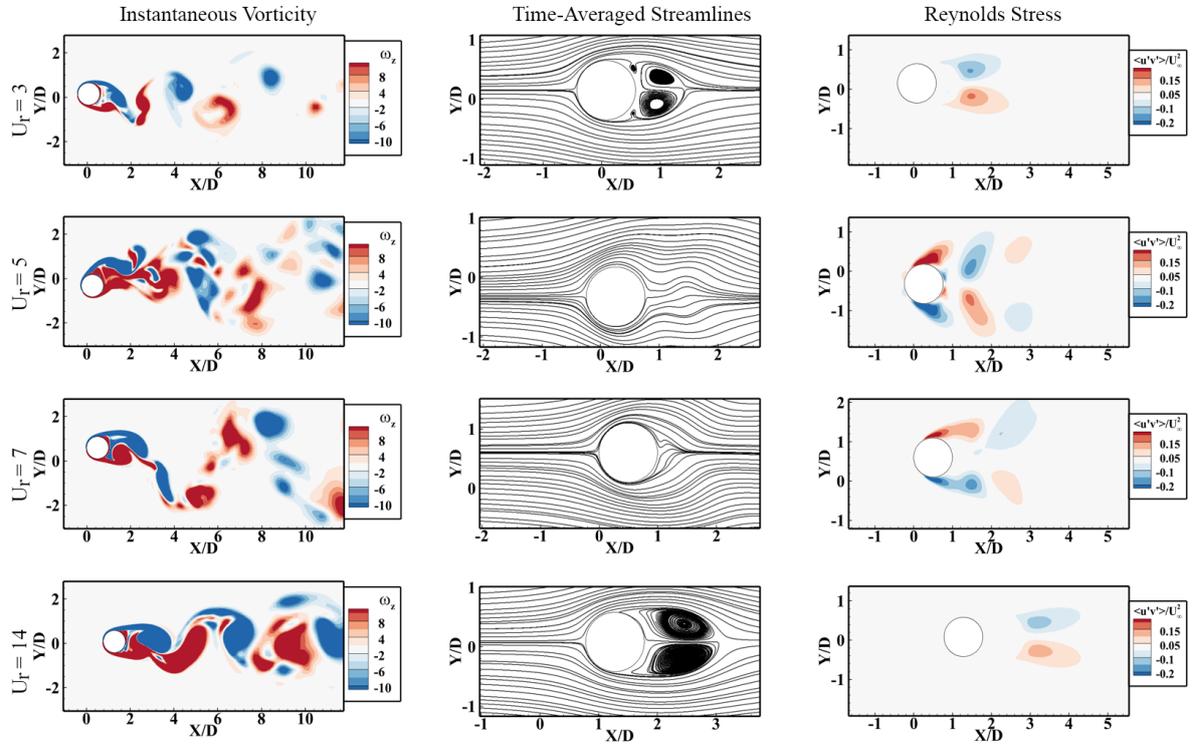

**FIG. 13.** Flow characteristics behind the oscillating cylinder under 2-DOF VIV system at Re = 1000 at different reduced velocities. [ $m^* = 2.546, \zeta = 0, f^* = 1.0, \text{Re} = 1000$ ]

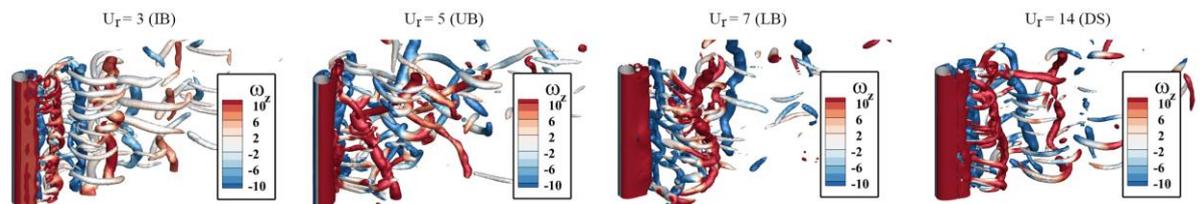

**FIG. 14.** Iso-surface of non-dimensional eigen value ($e_2$) and the contours of the spanwise vorticity (on the iso-surfaces) behind the 2-DOF system at Re = 1000 at different reduced velocities. [ $m^* = 2.546, \zeta = 0, f^* = 1.0, \text{Re} = 1000$ ]

## C. Wake Flow Transition for Cylinder Undergoing 2-DOF VIV

According to the preceding discussion, the wake for Re = 150 is two-dimensional, while at Re = 1000, it is fully three-dimensional. Thus, this transition in the wake from fully 2D to fully 3D takes place between the Re = 150 and 1000 for the 2-DOF VIV system. Here, in this section, we have investigated the 2D to 3D transition of the cylinder wake at a constant reduced velocity $U_r = 6$ (which lies in the lock-in regime for the range of $150 \leq \text{Re} \leq 1000$) with the increase in the Reynolds number ($\Delta \text{Re} = 50$). This section is divided into many subsections, where the first two sub-sections deal with the flow-wake visualization and frequency-amplitude response of the oscillations, respectively. The other subsections highlight the effect of different VIV parameters on the wake transition (if any), such as mass-ratio ($m^*$), damping ratio ($\zeta$), and frequency ratio ($f^*$).

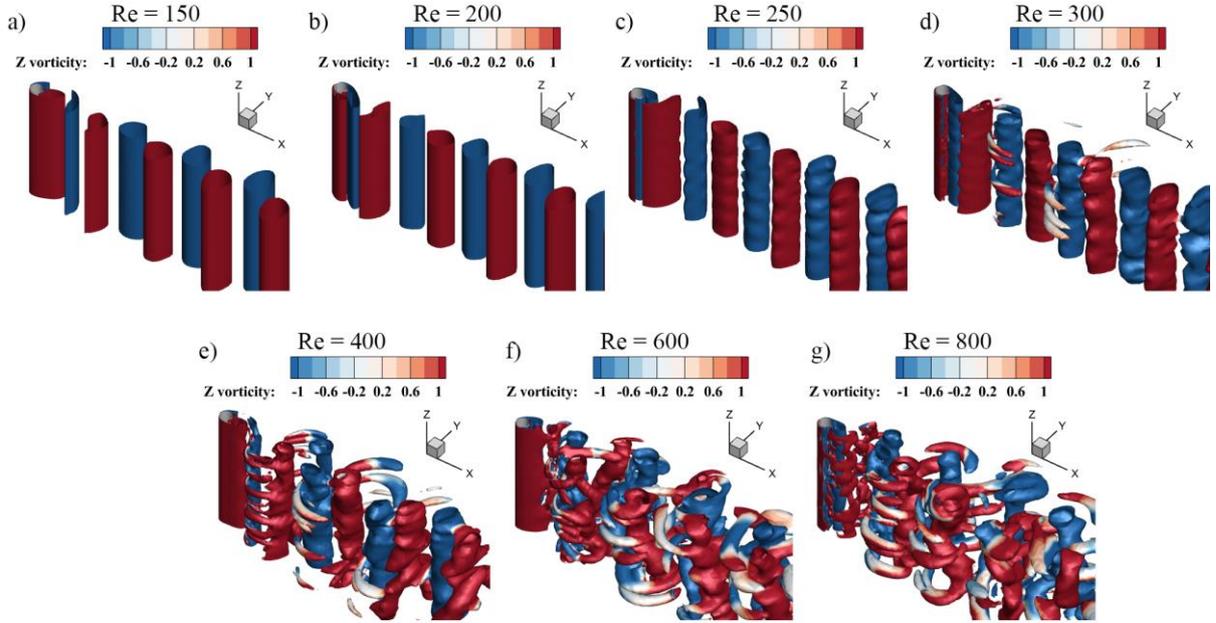

**FIG. 15.** Iso-surfaces of the eigenvalue $e_2 = 0.01$ coloured with the z-vorticity (plotted over the iso-surfaces) for a reduced value of $Ur = 6$ for different Reynolds numbers: (a) Re = 150, (b) Re = 200, (c) Re = 250, (d) Re = 300, (e) Re = 400, (f) Re = 600, and (g) Re = 800. [ $m^* = 2.546, \zeta = 0, f^* = 1.0, U_r = 6$ ]

### C.1. Wake-Flow Visualization

Figure 15 shows the iso-surface of the non-dimensional eigenvalue ($e_2 = 0.01$) colored with the instantaneous spanwise-vorticity at different Reynolds numbers. The cylinder sheds one pair of vortices in one period of oscillation. Miller and Williamson[60] reported the highest critical Reynolds number for the transition of the wake behind a stationary cylinder to be about 200, while for the transversely oscillating cylinder (1-DOF system), the transition from 2D to 3D takes place around Re = 250-300[17]. For a 2-DOF system, the vortex shedding is purely two-dimensional for Re < 250. At Re = 250, we observe the introduction of the three-dimensionality in the form of wavy spanwise vortices, which is seen in the iso-surfaces in Fig. 15.

Williamson[17] classified the two-different modes of transition to three-dimensionality (i.e., *mode-A* and *mode-B*) based on the wavelength of such spanwise vortices for a stationary cylinder. He reported the *mode-A* transition occurrence at the early stage and can be characterized by the wavelength of the spanwise vortices to be between 3 to 4 times the cylinder diameter[38, 61]. The wake flow transitions to *mode-B* with the increase in Reynolds number, and *mode-B* can be characterized by the spanwise wavelength of about one cylinder diameter. The transversely oscillating cylinder (1-DOF VIV system) exhibits the higher value of the critical Reynolds number for wake-transitioning from two-dimensional to three-dimensional[20]. Leontini et al.[8] observed that at higher oscillation amplitudes, the critical Reynolds number for mode-A exceeds the critical Reynolds number of *mode-B*, and *mode-B* occurs before *mode-A* for the transversely oscillating cylinder. Thus, the transition from 2D to 3D in the 1-DOF VIV system occurs directly with *mode-B*, and *mode-A* is absent[20]. Figure 16 portrays the iso-surfaces of the axial vorticity ($\omega_x$) along with the spanwise vorticity ($\omega_z$) for the 2-DOF VIV system (with $f^* = 1$). Figure 16 highlights that the wake flow Re < 250 is completely two-dimensional and governed by the spanwise vortices. Further increase in Re results in the appearance of the pairs of the streamwise vortices, whose dominance gets stronger with the increase in the Reynolds number.

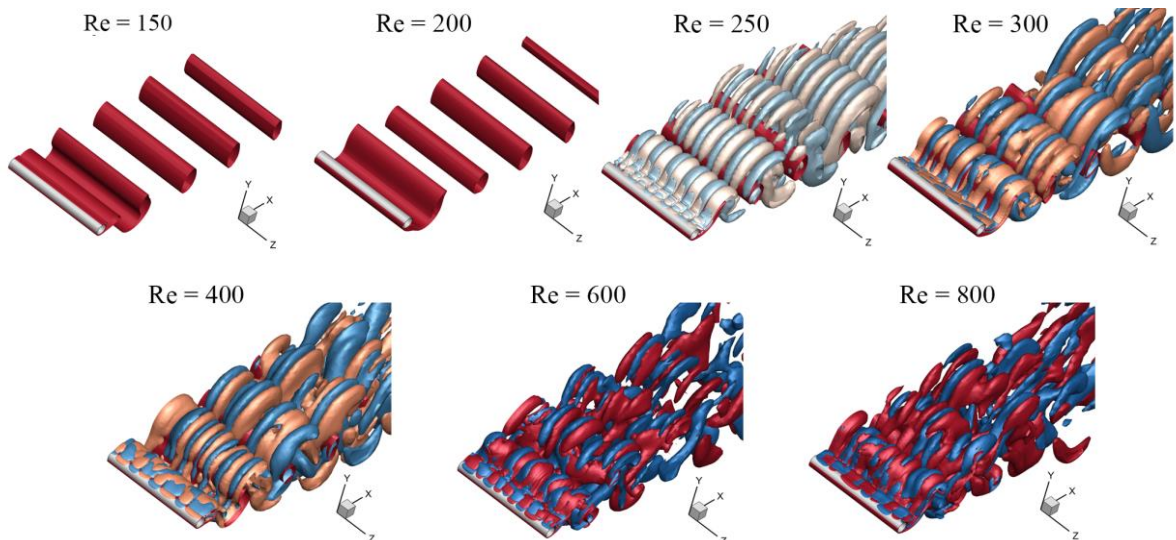

**FIG. 16.** Simulation results for 2-DOF VIV at different Reynolds number (*Mode-C* 3-D instability has been reported for the first time for the 2-DOF VIV system). The surfaces colored pink and blue mark a particular positive and negative value of the streamwise vorticity ($\omega_x$), and the red surface marks the value of spanwise vorticity ($\omega_z$). [$m^* = 2.546, \zeta = 0, f^* = 1.0, U_r = 6$]

Thus, the wake transition from 2D to 3D occurs at Re = 250, and the number of the streamwise pairs is equal to the number of spanwise waves in the iso-surfaces of the eigenvalue, $e_2$, resulting in the spanwise wavelength of 1.92 times the cylinder diameter. Compared with the stationary cylinder modes, the instability mode it follows for 2D to 3D transition depends on the number of streamwise vortex pairs (or on the wavelength of the spanwise vortices). The wavelength of spanwise vortices ($\lambda_z / D \approx 1.92$) belongs to neither *mode-A* ($\lambda_z / D \approx 3-5$) nor *mode-B* ($\lambda_z / D \approx 1.0$). It lies between both modes and resembles the *mode-C* ($\lambda_z / D \approx 1.5-2.5$), which is being reported for the first time in the literature on the 2-DOF VIV system. The vortex flow pattern at a higher Reynolds number is similar to Re = 300 but less regular. The streamwise pairs of the vortices get stronger with an increase in the Re, resulting in the dislocation of the spanwise vortices. Thus, at higher Reynolds numbers, the streamwise vortices dominate the wake, and the spanwise vortices are very irregular, as seen in the iso-surfaces of the spanwise vorticity in Fig. 17.

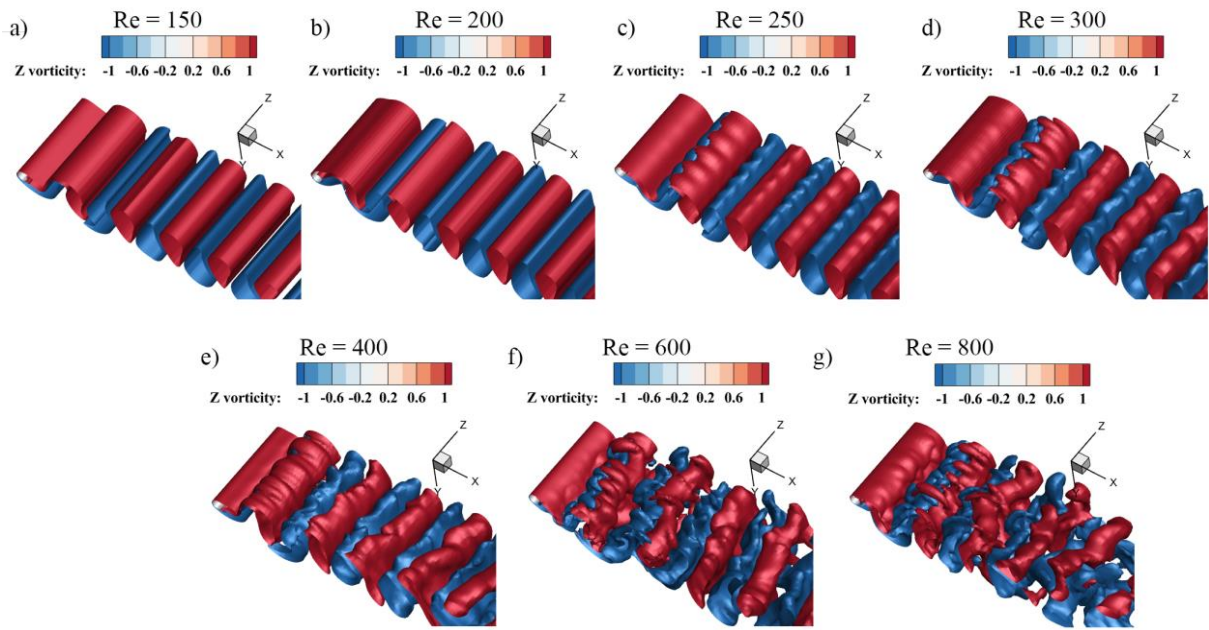

**FIG. 17.** Iso-surfaces of the spanwise Vorticity for the cylinder under 2-DOF VIV at different Reynolds numbers: (a) Re = 150, (b) Re = 200, (c) Re = 250, (d) Re = 300, (e) Re = 400, (f) Re = 600, and (g) Re = 800. [ $m^* = 2.546, \zeta = 0, f^* = 1.0, U_r = 6$ ]

Further, we look at the vortex shedding modes behind the cylinder at different Reynolds numbers via instantaneous spanwise vorticity at three different positions along the cylinder span, i.e., at z/L=0, 0.5, and 1.0. The slices are reported for four Re values, Re = 150, 300, 400, and 600 in Fig. 18. The cylinder at Re = 150 exhibits the same vortex-shedding (2S type) throughout its span (Fig. 18 (a)), which is the typical characteristics of a fully two-dimensional wake. For Re = 300 (which lies in the transition regime, Fig. 18 (b)), the cylinder sheds the vortices in the 2S type of shedding, and a little difference is observed at different spanwise locations due to weak three-dimensionality. At a higher Reynolds number (Re = 400), the shedding converts to the 2P type of shedding, where the cylinder sheds two pairs of the vortices in one complete oscillation cycle (as seen in Fig. 18 (c)). The vortex shedding is significantly different at each spanwise location due to the wrinkling of the spanwise vortex tubes at a higher Re value (as seen in Fig. 18 (d)).

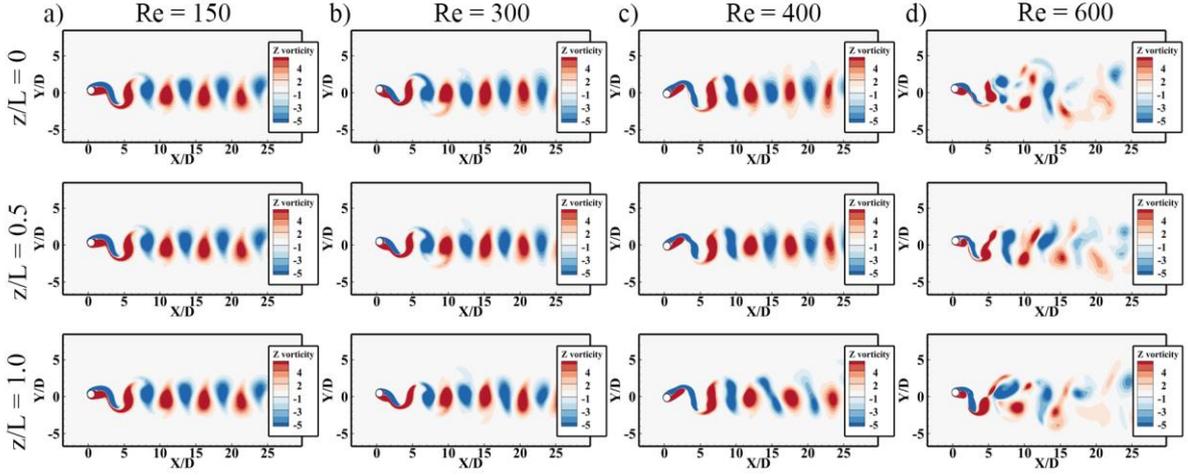

**FIG. 18.** Instantaneous spanwise vorticity contours at three different spanwise-locations, i.e., z/L = 0, 0.5, and 1.0 at: (a) Re = 150, (b) Re = 300, (c) Re = 400, and (d) Re = 600. [ $m^* = 2.546, \zeta = 0, f^* = 1.0, U_r = 6$ ]

Figure 19 shows the pictures of the instantaneous streamwise vorticity on the x-z plane at the cylinder's center for different Re. It can be seen that the quantities of the Re = 150 and 200 are significantly lower than the ones of higher Re. This reconfirms the dominance of the streamwise vorticity during the transition. The approximate wavelengths (calculated based on the appearance of the alternate pairs in Fig. 19) for the Re > 250 are 1.92 times the cylinder's diameter. This also indicates the transition to take place via *mode-C*.

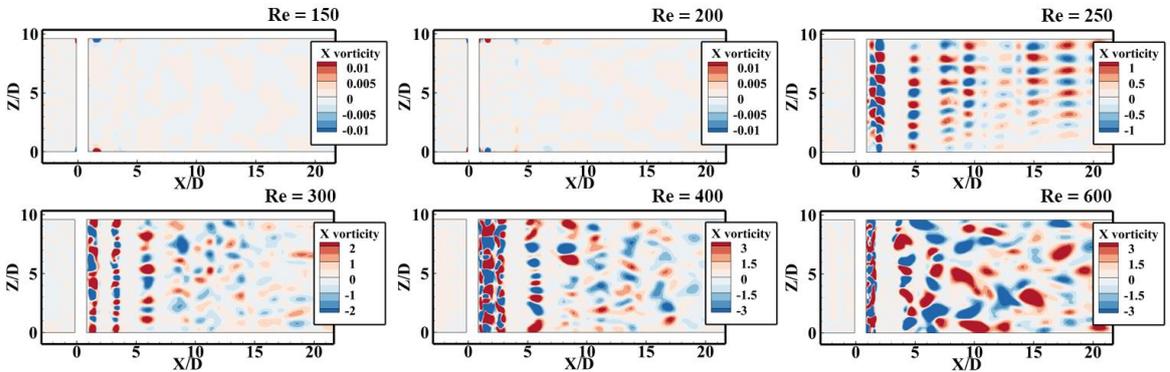

**FIG. 19.** Instantaneous streamwise vorticity on the x-z plane at the cylinder's center. [ $m^* = 2.546, \zeta = 0, f^* = 1.0, U_r = 6$ ]

To identify the mode via which the 3D instability sets in, the presence/absence of period doubling is studied for the Re = 250 wake flow. If T is the time period of the vortex shedding behind the cylinder, Fig. 20 shows the instantaneous streamwise vorticity for the Re = 250 flow at three-time instants corresponding to t, t+T, and t+2T. We observe that $\omega_x(t) = -\omega_x(t+T)$, and $\omega_x(t) = \omega_x(t+2T)$ i.e., the flow has a time period of 2T. Further, Fig. 21 (a) and (b) display the spanwise variation of the streamwise vorticity in the wake at (x/D, y/D) = (5, 1.7) for Re = 200, and 250, respectively, for the three-time instants. One can observe a very low value of the streamwise vorticity at Re = 200 due to the two-dimensional wake governed by the spanwise vorticity. For the three-dimensional wake at Re = 250, the streamwise vorticity is out of phase for 't+T' time instant relative to the vorticity at time instant 't' and comes to in-phase at 't+2T'. The out-of-phase/in-phase behavior of the streamwise vorticity in Fig. 21 (b) shows the alternate occurrence of the streamwise vortices with the positive and negative values in the wake at different time instants 't', 't+T', and 't+2T'. The distribution of these streamwise vortices in the cylinder wake determines the mode it follows for the three-dimensional transition. The number of peaks at any time instant in Fig. 21 (b) represents the presence of the alternate streamwise vortices pair. From these variations of the streamwise vorticity over the cylinder span, the spanwise wavelength of the three-dimensional instabilities in the flow is estimated to be 1.92, which re-confirms that the flow is associated with *mode-C* instability.

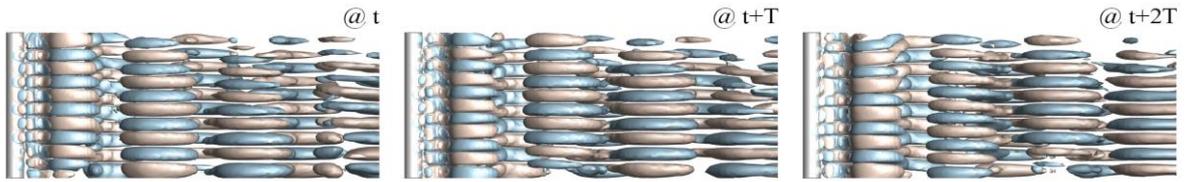

**FIG. 20.** Iso-surfaces of streamwise component of vorticity ($\omega_x = \pm 0.4$) at Re = 250 to illustrate the *mode-C* instability. [ $m^* = 2.546, \zeta = 0, f^* = 1.0, U_r = 6$ ]

We further investigate the RT symmetry (R: reflection about the wake axis, T: translation in time) in the wake flow (if any) at Re = 250. Fig. 21 (c) shows the space-time variation of streamwise vorticity in the wake along the y-axis at (x/D, z/D) = (5, 4.8) for a few cycles of primary vortex shedding. This space-time variation is not symmetric about the wake axis and shows variation with the translation in time. Thus, we can conclude that the wake flow exhibits neither odd-RT nor even-RT symmetry at the Re = 250 for a 2-DOF VIV system.

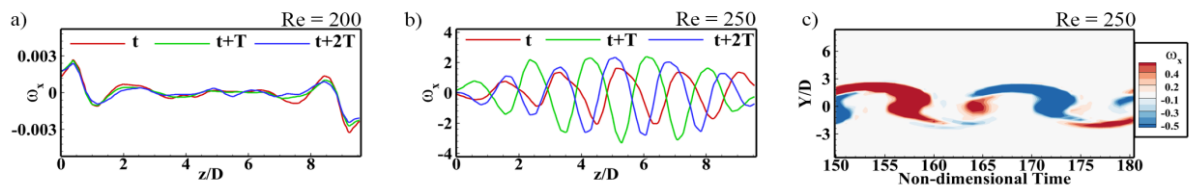

**FIG. 21.** Spanwise variation of $\omega_x$ in the wake at (x/D, y/D) = (5, 1.7) for (a) Re = 200, and (b) Re = 250; (c) Time history of streamwise vorticity along the y-axis at (x/D, z/D) = (5, 4.8) [ $m^* = 2.546, \zeta = 0, f^* = 1.0, U_r = 6$ ]

## C.2. *Force, Oscillation Amplitude, and Frequency Response*

Figure 22 represents the oscillation amplitude response along with the corresponding aerodynamic forces and the frequency response with the variation in the Reynolds number. The aerodynamic lift coefficient shows the r.m.s. value of the surface averaged lift coefficient, while the drag coefficient shows the mean of the surface averaged drag coefficient. For the fully two-dimensional wake, the lift and drag coefficients drop with the increase in the Re and the increase in the oscillation frequency in streamwise and transverse directions. The oscillation amplitude remains almost constant even with the variation in the lift coefficient till the transition at Re = 250, as seen in Fig. 22 (a) and (b). With the appearance of the streamwise vortices pairs, the flow behind the cylinder transitions to the three-dimensional resulting in the continuous increase in the streamwise and the transverse oscillation amplitudes. The oscillation frequencies drop during the transition and then increase linearly with the Re, as seen in Fig. 22 (c).

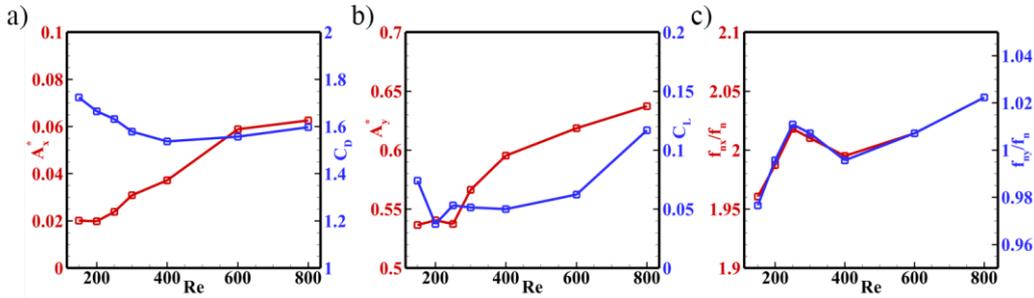

**FIG. 22.** Aerodynamic performance of the cylinder oscillating under 2-DOF VIV system: (a) the streamwise mean oscillation amplitude and its corresponding mean drag coefficient, (b) the transverse mean oscillation amplitude response, and the r.m.s. value of the lift coefficient on the cylinder's surface, (c) frequency response over the range of Re [ $m^* = 2.546, \zeta = 0, f^* = 1.0, U_r = 6$ ]

Figure 23 displays the time history of the streamwise and transverse oscillation amplitude, their corresponding lift, drag force coefficients, and their frequency spectrum. The frequency spectrum highlights the role of transition with the appearance of the dominant higher frequencies. For the low Re flows (till the wake is entirely two-dimensional), the frequency spectrum shows the dominance of the single peak frequency for streamwise and transverse oscillations. Also, these peak frequency matches the corresponding force coefficient's frequency. With the transition to three-dimensionality, the number of harmonics starts to appear and dominate the cylinder oscillations. Under the influence of these peak harmonics, the amplitude response loses its periodic nature and becomes non-periodic, as seen in the transverse oscillation amplitude response in Fig. 23 (c).

Further, to comprehend the effect of the flow transition on the forces and the spatiotemporal nature of the wake flow, the contours of the sectional average pressure coefficient have been plotted on the z-t plane for $U_r = 6$ and at different Reynolds numbers in Fig. 24. The sectional average pressure coefficient [$C_{P_{avg}}(z,t)$], at any spanwise location, represents the integrated signature of the flow past the cylinder at that spanwise section. The pressure coefficient along the span is constant for Reynolds numbers, where the flow is entirely two-dimensional ($Re < 250$). As shown in Fig. 15 and 16, the three-dimensionality appears for the first time at a Reynolds number of 250, resulting in a variation of the average pressure along the cylinder span as seen in Fig. 24. As a portion of the spanwise vortices' energy is transferred to the streamwise vortices which are perpendicular to the cylinder, they no longer contribute to the lift force and results in the reduced pressure values over the cylinder at the higher Reynolds numbers. Also, the sectional average pressure coefficient variation over the cylinder span gets stronger with the increase in the Re values.

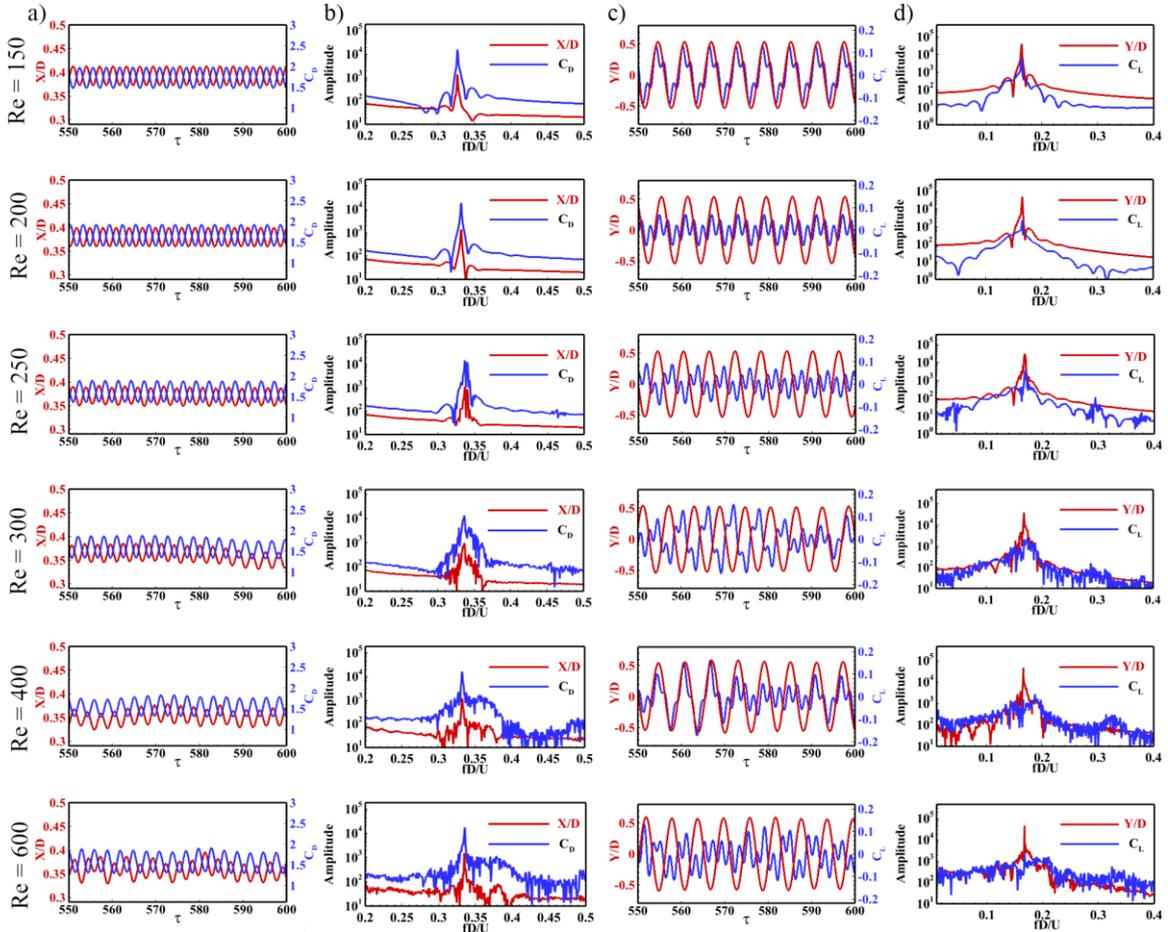

**FIG. 23.** Time history of amplitude and frequency response of the circular cylinder undergoing the 2-DOF VIV oscillations: (a) streamwise oscillation amplitude and the aerodynamic drag coefficient, (b) the frequency spectrum of the streamwise oscillation amplitude and the drag coefficient, (c) the transverse oscillation amplitude and the lift coefficient, and (d) the frequency spectrum of the transverse oscillation amplitude and the lift coefficient [$m^* = 2.546, \zeta = 0, f^* = 1.0, U_r = 6$]

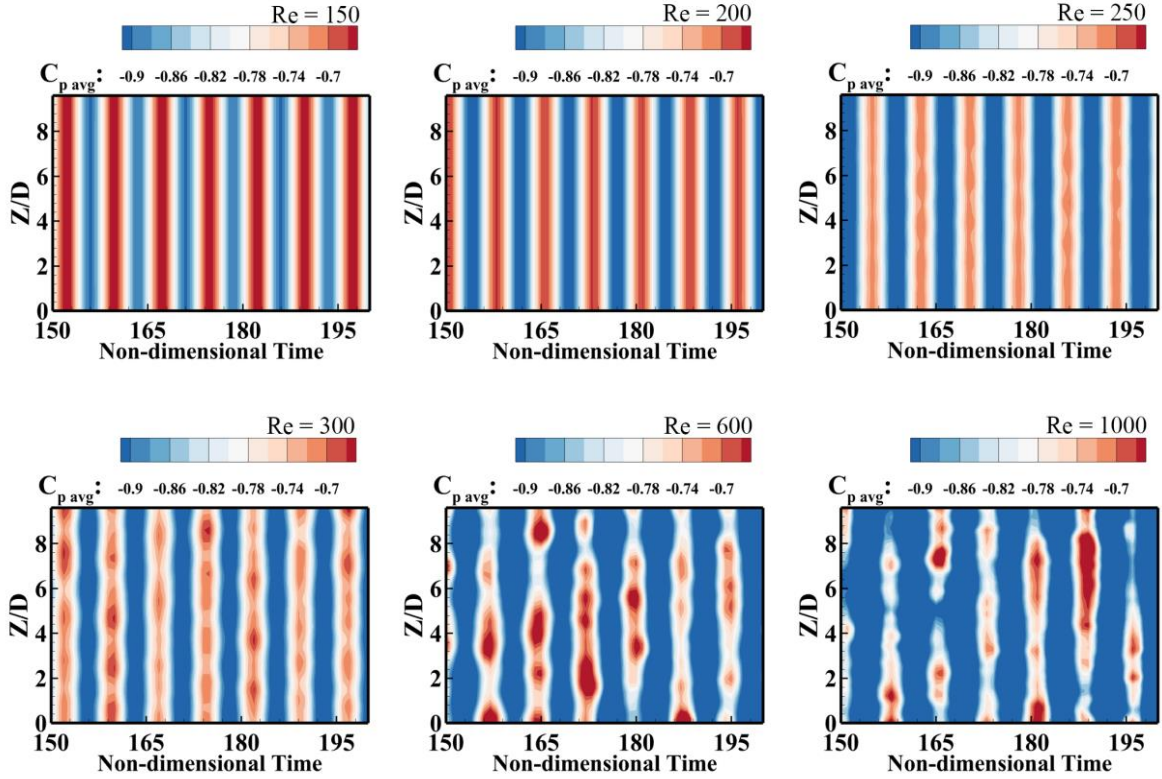

**FIG. 24.** Contours of sectional average pressure coefficient on the z-t plane for different Reynolds numbers. [ $m^* = 2.546, \zeta = 0, f^* = 1.0, U_r = 6$ ]

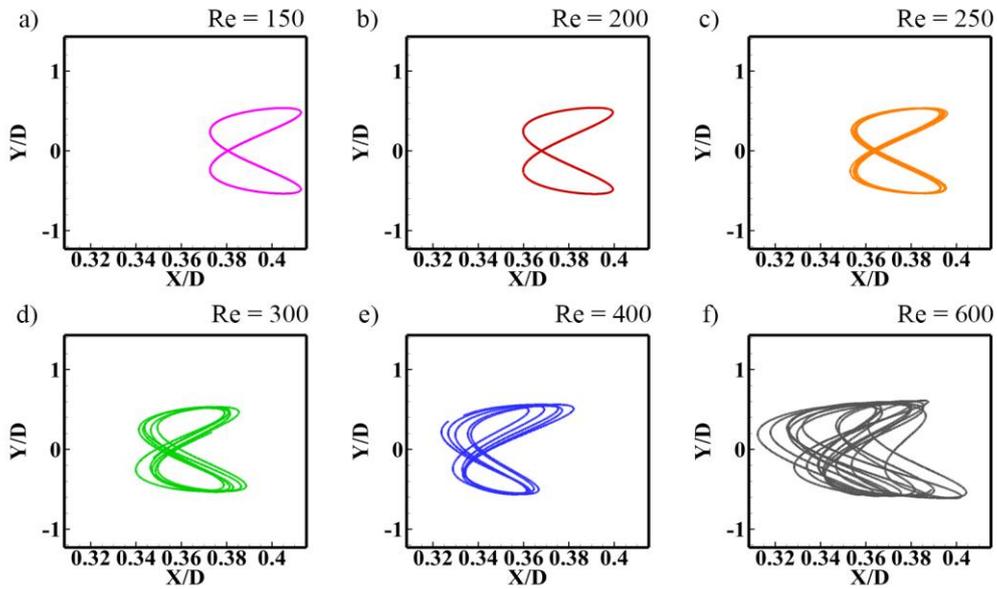

**FIG. 25.** Phase portraits between the streamwise and the transverse oscillations of the cylinder undergoing 2-DOF VIV at different Reynolds numbers: (a) Re=150, (b) Re=200, (c) Re=250, (d) Re=300, (e) Re=400, (f) Re=600 [ $m^* = 2.546, \zeta = 0, f^* = 1.0, U_r = 6$ ]

Fig. 25 portrays the phase portraits between the streamwise oscillation amplitude and the transverse oscillation amplitude for a few oscillation cycles to understand the effect of wake transition on the cylinder's motion trajectory. The cylinder traces the eight-shaped trajectory during its motion under 2-DOF. For the fully two-dimensional wake ( Re < 250 ), the cylinder follows the same trajectory during each cycle. With the introduction of the three-dimensionality in the wake at Re = 250, the cylinder oscillations lose their periodic nature (as discussed in the previous subsections), and the cylinder no longer follows the same trajectory for each cycle. This is more dominantly visible for the higher Re values, as reported in Fig. 25.

*C.3. Effect of Mass Ratio*

Govardhan and Williamson[18] observed that the results for the low and high mass ratios differ. The oscillation amplitude decreases with the mass ratio increase, and the wake strongly depends on the oscillation amplitude and frequency. Thus, to investigate the effects of mass ratio on the cylinder's wake and its transition from 2D to 3D, we have performed simulations for three different mass ratios, i.e., $m^* = 2.546$, 10, and 70. Reynolds number varies from 150 to 1000, keeping other simulation parameters fixed at $\zeta = 0, f^* = 1.0$, and $U_r = 6$. The non-dimensional eigenvalue ($e_2 = 0.01$) is used to identify the onset of three-dimensionality for different mass ratios. Figure 26 shows that the wake flow is two-dimensional $Re < 300$, and the onset of the three-dimensionality occurs first at $Re = 300$ for $m^* = 10$ (see appendix A.2). While, for the higher mass ratio ($m^* = 70$), the wake flow transition occurs as early as $Re = 200$. This early transition can be attributed to the lower amplitude and higher frequency oscillations of the cylinder at the higher mass ratios, as depicted in the oscillation amplitude and frequency response in Fig. 27. The onset of the three-dimensionality for higher mas ratios occurs near the critical Re for the fixed rigid cylinder[17].

Based on the spanwise wavelength from Fig. 26, the mode it takes for the onset of the three-dimensional instability can be identified. The spanwise wavelength for $m^* = 10$ is calculated to be 1.92*D, which corresponds to the *mode-C* transition, while for $m^* = 70$, the spanwise wavelength is found to be 3.20*D, indicating the 2D to 3D wake transition occurring via *mode-A* instability. Further, to reconfirm these instability modes, the iso-surfaces of the streamwise vorticity and the spanwise vorticity are plotted in Fig. 28 for three different mass ratios at the critical Reynolds number corresponding to the respective mass ratio. One can identify five pairs of the streamwise vortices for $m^* = 10$ and three pairs for $m^* = 70$. Thus, the calculated wavelength (based on the number of the streamwise vortices) is found to be 1.92*D and 3.20*D for the m* = 10 and 70, respectively. This matches well with the spanwise wavelength and confirms the *mode-C* and *mode-A* transition for m* = 10 and 70, respectively.

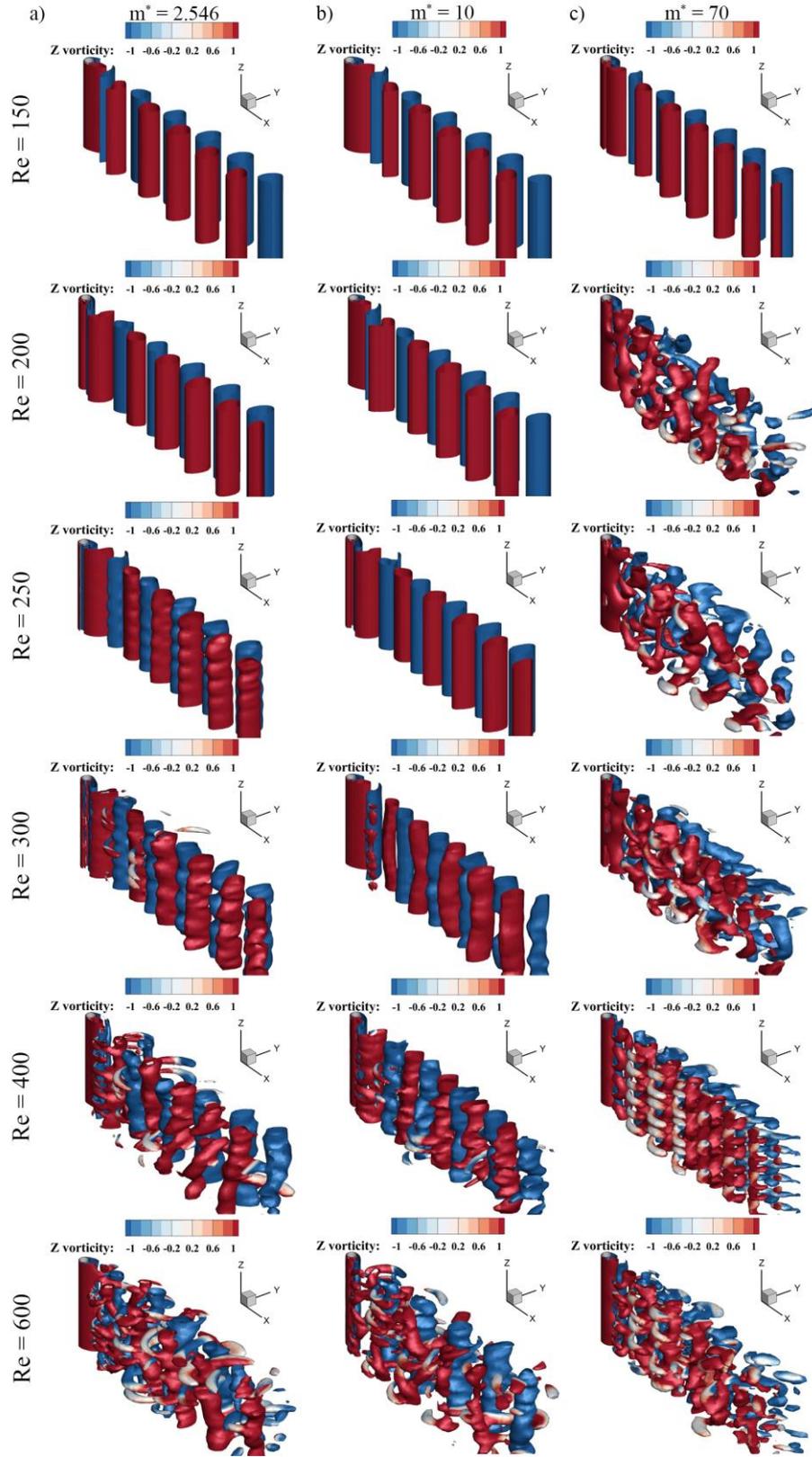

**FIG. 26.** Iso-surfaces of the eigenvalue $e_2 = 0.01$ colored with the z-velocity (plotted over the iso-surfaces) for a reduced value of $U_r = 6$ for different Reynolds numbers ($150 \leq \mathrm{Re} \leq 600$) for different mass ratios: (a) $m^* = 2.546$, (b) $m^* = 10$, and (c) $m^* = 70$ [$\zeta = 0, f^* = 1.0, U_r = 6$]

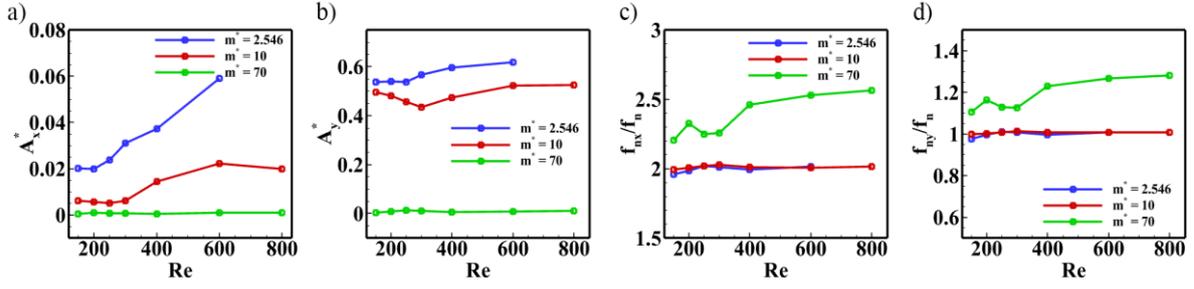

**FIG. 27.** Oscillation amplitude and the frequency response for the cylinder at different mass ratios for variation of Re: (a) streamwise mean oscillation amplitude, (b) transverse mean oscillation amplitude, (c) streamwise oscillation frequency normalized with the natural streamwise frequency of the cylinder, and (d) transverse oscillation frequency normalized with the natural transverse frequency of the cylinder [ $\zeta = 0, f^* = 1.0, U_r = 6$ ]

It is interesting to note here that the cylinder with m* = 70 takes *mode-A* for 2D to 3D transition of the wake, which resembles the case of the stationary rigid cylinder. Thus, we can conclude that the oscillation amplitude for the m* = 70 is very low, and the cylinder almost acts as a stationary cylinder for the mentioned VIV parameters.

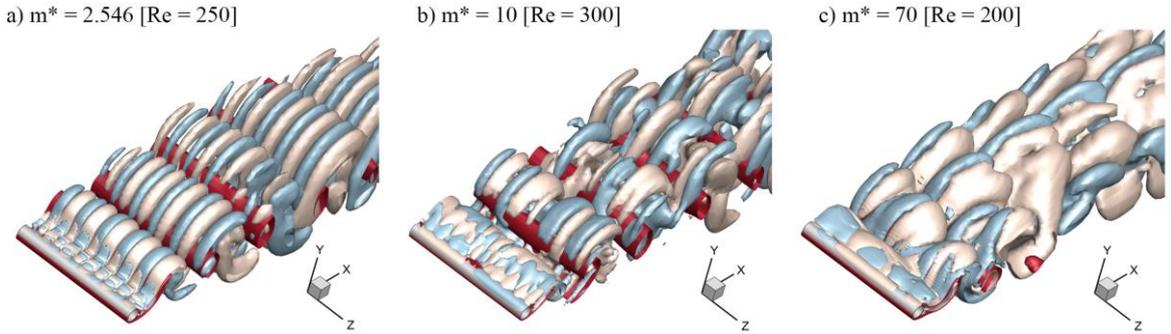

**FIG. 28.** The iso-surfaces (colored pink and blue) mark a particular positive and negative value of the streamwise vorticity ( $\omega_x$ ), and the red surface marks the value of spanwise vorticity ( $\omega_z$ ) for different mass ratios: (a) $m^* = 2.546$, (b) $m^* = 10$, and (c) $m^* = 70$ [ $\zeta = 0, f^* = 1.0, U_r = 6$ ]

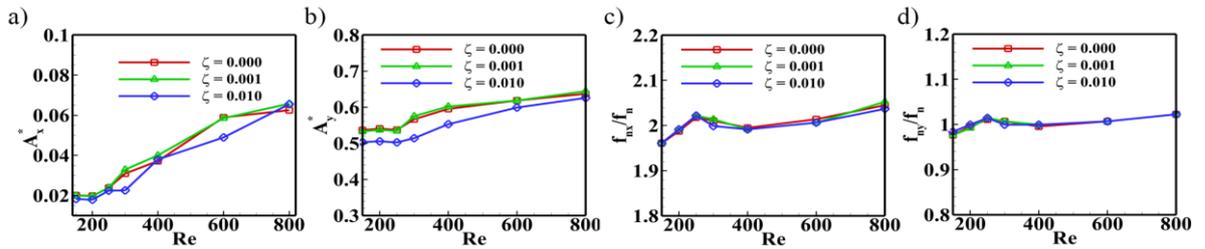

**FIG. 29.** Effect of the damping ratio on the amplitude and frequency response of the circular cylinder over the range of Re: (a) mean streamwise oscillation amplitude, (b) mean transverse oscillation amplitude, (c) normalized streamwise oscillation frequency, and (d) normalized transverse oscillation frequency [ $m^* = 2.546, f^* = 1.0, U_r = 6$ ]

## C.4. Effect of Damping Ratio

Further, the simulations are performed to confirm the effect of another critical VIV parameter, damping ratio ($\zeta$), on the 2D to 3D transition. Five different damping ratio values ranging from 0.0 (corresponding to the undamped system), to 1.0 (corresponding to the highly damped system) are compared for a cylinder with a mass ratio of 2.546. Fig. 29 shows the oscillation amplitude and the frequency response over the range of Re for three different damping ratios. Increasing the damping of the system opposes the motion of the cylinder. It results in reduced oscillation amplitudes in both directions at Re values (as seen in Fig. 29(a) and (b)). From the frequency response of Fig. 29 (c) and (d), one can observe that the damping ratio barely affects the oscillation frequency, and the cylinder oscillates with the same frequency for all the damping ratio values. Also, the reduction in amplitude is not that significant, and the transition of the wake flow from 2D to 3D takes place at a similar Reynolds number, i.e., $200 \leq \text{Re} \leq 250$ for all the damping ratios. Further, Fig. 30 shows the iso-surfaces of the axial vorticity and the spanwise vorticity to identify the transition mode. Around five to six pairs of streamwise vorticity exist over the cylinder span, indicating the spanwise wavelength in the range of 2.13*D - 1.61*D. This confirms that the 2D to 3D transition occurs via *mode-C* for all the damping ratios.

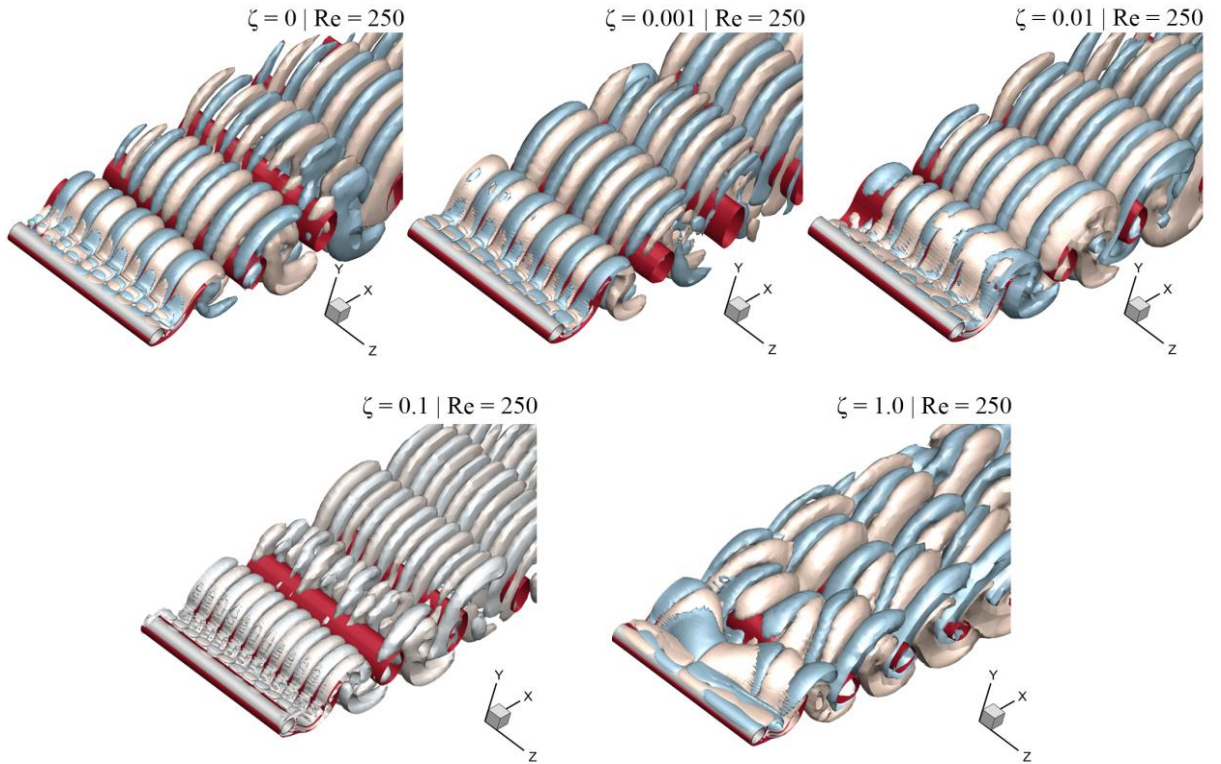

**FIG. 30.** Effect of the damping ratio ($\zeta$) on the vorticity behind the cylinder for 2-DOF VIV.
[ $m^* = 2.546, f^* = 1.0, U_r = 6$ ]

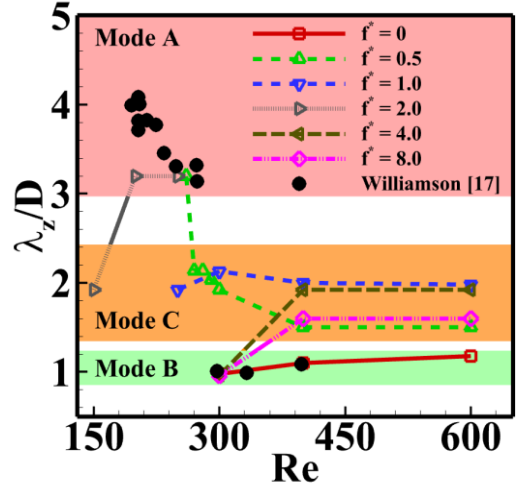

**FIG. 31.** Spanwise instability wavelength of the three different 3-D instabilities (*mode-A, B,* and *C*) normalized with the cylinder diameter ($\lambda_z / D$) for the 2-DOF VIV system. [$m^* = 2.546, \zeta = 0, U_r = 6$]

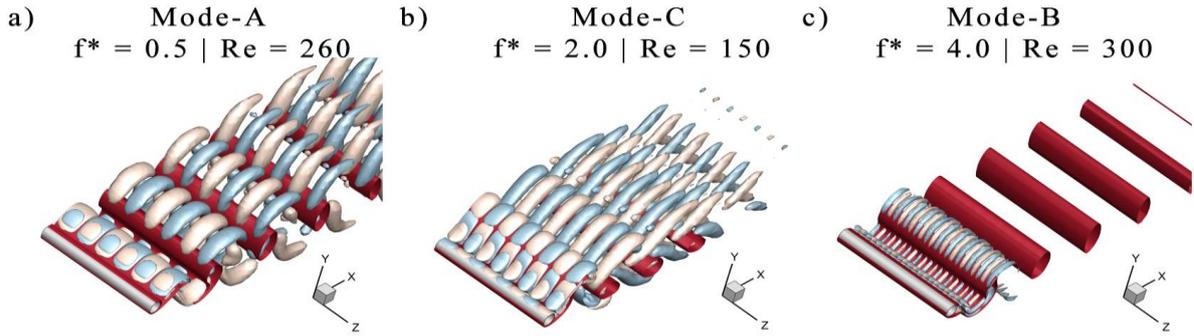

**FIG. 32.** Simulation results for *Mode-A, B*, and *C* 3-D instabilities. The existence of *Mode-C* 3-D instability has been reported for the first time for the 2-DOF VIV system. The surfaces colored pink and blue mark a particular positive and negative value of the streamwise vorticity ($\omega_x$), and the red surface marks the value of spanwise vorticity ($\omega_z$): (a) at $f^* = 0.5$, Re = 260, (b) $f^* = 2.0$, Re = 150, and (c) $f^* = 4.0$, Re = 300 [$m^* = 2.546, \zeta = 0, U_r = 6$]

## C.5. Effect of Frequency Ratio

One of the objectives of this study is to look at the effect of the natural frequency ratio ($f^*$) on the vortex shedding and the three-dimensionality in the cylinder wake for 2-DOF VIV. The frequency ratio is an important parameter as it governs the trajectory of the cylinder's oscillation. The simulations also consider the frequency ratio from 0 to 8 and report the three-dimensionality effects on the wake corresponding to different frequency ratios. Figure 31 presents the $\lambda_z - \text{Re}$ map for the 2-DOF VIV of a circular cylinder at a low mass ratio (m* = 2.546). The map is compared with the data obtained by Williamson[14] for the stationary cylinder. Different modes of three-dimensional instabilities are reported for different frequency ratios. The $f^* = 0$ case represents the case of 1-DOF VIV in the transverse direction. The transition from 2D to 3D wake occurs at a Re = 300 via *mode-B* instability (see appendix A.3), which agrees with the previous literature on the wake three-dimensionality for 1-DOF VIV system[8, 10, 20]. When the frequency of the streamwise oscillation is half of the transverse oscillation frequency ($f^* = 0.5$), the transition takes place at Re = 260 with a *mode-A* instability followed by the *mode-C* instability with the increase in the Reynolds number. The case of interest is $f^* = 2$ when the streamwise oscillation frequency is twice the transverse oscillation frequency. Here, the transition occurs at a very low Reynolds number (Re = 150) via *mode-C* instability, which transitions to *mode-A* with the increase in the Reynolds number. Beyond Re = 260, due to the high oscillation amplitude in streamwise and transverse directions, the three-dimensionality was so strong that no-identifiable modes were visible. Further, for higher frequency ratios ($f^* = 4, 8$), the transition occurs at Re = 300, with the *mode-B*, followed by the *mode-C* with the increase in the Re value. Thus, the *mode-C* instability is the preferred one for the 2-DOF VIV as opposed to the *mode-B* (for 1-DOF VIV). Figure 32 portrays the *mode-A, B,* and *C* instabilities obtained for the different combinations of the frequency ratio and the Reynold number at the onset of the wake three-dimensionality.

## V. CONCLUSION

The present study explores the wake transition (2D to 3D) characteristics of an elastically mounted circular cylinder under 2-DOF VIV. The effect of various VIV parameters, such as Reynolds number, reduced velocities, mass ratios, damping ratios, and frequency ratios, are widely investigated by employing the wake-flow visualization, aerodynamic force coefficients, mean oscillation amplitude, and frequency response. The iso-surfaces of the non-dimensional eigenvalue ($e_2 = -\lambda_2 / (U/D)^2$) of the tensor $S^2 + \Omega^2$ are utilized to visualize the three-dimensional vortices behind the cylinder. The cylinder exhibits two branch amplitude responses at lower Re values: the initial branch (IB) and the lower branch (LB). At higher Re values, another high-oscillation-amplitude branch, upper branch (UB), is obtained. No 'super-upper' branch is obtained at Re = 1000 for the mass ratio $m^* = 2.546$, as reported by Jauvtis and Williamson[23] for such low mass ratios. This may be because of the variation in the Reynolds number ($2,000 \leq \text{Re} \leq 11,000$) during the experiments, while we have simulated at the constant Re of 1000. Further, the results for the wake three-dimensionality reveal that the cylinder wake undergoes 2D to 3D transition via *mode-C* of three-dimensional instabilities with a spanwise wavelength of 1.92. The spanwise wavelength of the three-dimensionality is further confirmed with the number of pairs of the streamwise vortices and matches well. Also, the onset of the three-dimensionality occurs around the critical Reynolds number of 250. Further, this critical Reynolds number for the 2D to 3D transition depends on the mass and frequency ratios. Different transition modes can be obtained by varying the frequency ratio and the Re.

**DATA AVAILABILITY**

The data that support the findings of this study are available from the corresponding author upon reasonable request.

**ACKNOWLEDGMENTS**

The authors would like to acknowledge the National Supercomputing Mission (NSM) for providing the computational resources of 'PARAM Sanganak' at IIT Kanpur, which is implemented by C-DAC and supported by the Ministry of Electronics and Information Technology (MeitY) and Department of Science and Technology (DST), Government of India. The authors would also like to acknowledge the IIT-K Computer center (www.iitk.ac.in/cc) for providing the resources to perform the computation work.

# APPENDICES
## A.1. CALCULATION OF GRID-CONVERGENCE INDEX (GCI)

The grid-convergence study is performed by taking Grid-3 as the base grid and approximating the error in Grid-4 compared to Grid-3 from the Richardson error estimator, defined by,

$$E_1^{fine} = \frac{\varepsilon_{43}}{1 - r_{43}^0} \tag{A.1}$$

Error in Grid-2, compared to the solution of Grid-3, is approximated by the coarse-grid Richardson error estimator, which is defined as,

$$E_1^{coarse} = \frac{r^0 \varepsilon_{32}}{1 - r_{32}^0} \tag{A.2}$$

Where r is the grid refinement ratio between the two consecutive grids defined as,

$$r_{i+1,i} = \frac{h_{i+1}}{h_i}$$

The error ($\varepsilon$) is estimated from the solutions of two consecutive grids by,

$$\varepsilon_{i+1,i} = \frac{f_{i+1} - f_i}{f_i} \tag{A.3}$$

Grid-Convergence Index (GCI), which accounts for the uncertainty in the Richardson error estimator, is calculated for fine-grid and coarse-grid as given[53-54],

$$GCI_{fine} = F_s \mid E_1^{fine} \mid \qquad (A.4)$$

$$GCI_{coarse} = F_s \mid E_2^{coarse} \mid \qquad (A.5)$$

Further, to calculate the order of accuracy, $L_2$ norm of the errors between is grids is calculated as,

$$L_2 = \sqrt{\left(\sum_{i=1}^{N} \mid \varepsilon_{i+1,i} \mid^2 / N\right)} \qquad (A.6)$$

Where N is the total number of grid points considered for the calculation of $L_2$ norm.

## A.2. WAKE TRANSITION FROM TWO-DIMENSIONAL TO THREE-DIMENSIONAL WAKE FOR 2-DOF VIV SYSTEM [ $m^* = 10$ ]

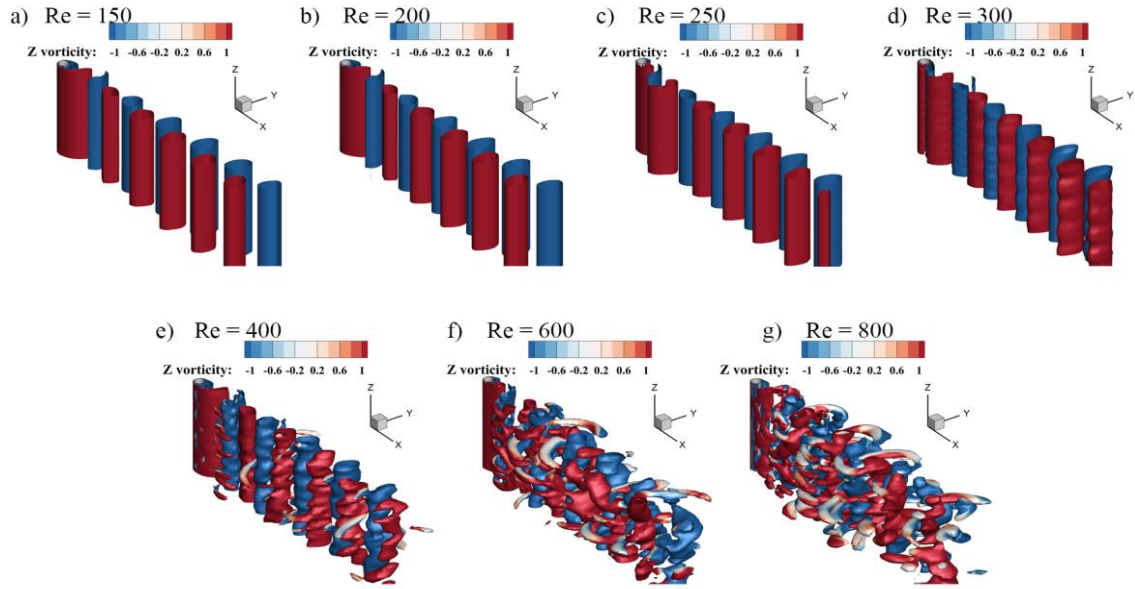

**FIG. 33.** Iso-surfaces of the non-dimensional eigenvalue $e_2 = 0.01$ coloured with the z-velocity (plotted over the iso-surfaces) for a reduced value of $Ur = 6$ for different Reynolds numbers: (a) Re = 150, (b) Re = 200, (c) Re = 250, (d) Re = 300, (e) Re = 400, (f) Re = 600, and (g) Re = 800. [ $m^* = 10, \zeta = 0, U_r = 6, 2-DOF$ ]

## A.3. WAKE TRANSITION FROM TWO-DIMENSIONAL TO THREE-DIMENSIONAL WAKE FOR 1-DOF VIV SYSTEM [ $m^* = 2.546$ ]

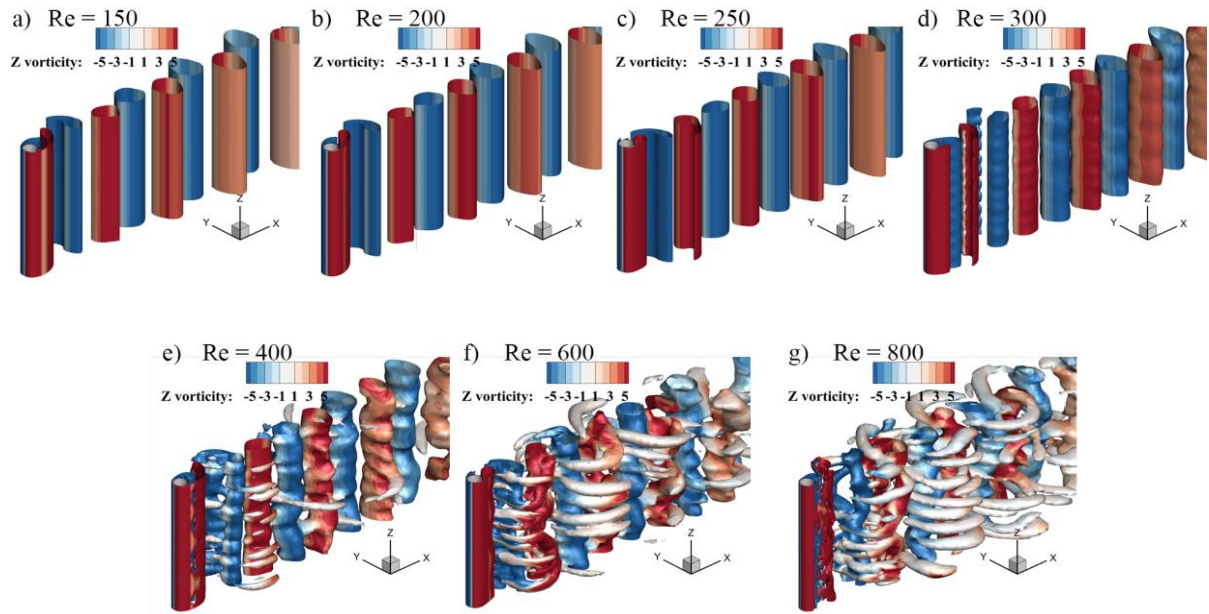

**FIG. 34.** Iso-surfaces of the non-dimensional eigenvalue $e_2 = 0.01$ coloured with the z-velocity (plotted over the iso-surfaces) for a reduced value of $Ur = 6$ for different Reynolds numbers: (a) Re = 150, (b) Re = 200, (c) Re = 250, (d) Re = 300, (e) Re = 400, (f) Re = 600, and (g) Re = 800. [ $m^* = 2.546, \zeta = 0, U_r = 6, 1 - DOF$ ]